\documentclass{llncs}
\pdfoutput=1
\usepackage[T1]{fontenc}
\usepackage{lmodern}
\usepackage[scaled]{}
\usepackage[english]{babel}
\usepackage[utf8]{inputenc}
\usepackage{cite}
\usepackage{listings}
\usepackage{latexsym, amssymb, amsmath, amsfonts, yfonts, mathrsfs, mathbbol}
\usepackage{syntax}
\usepackage{mathtools}
\usepackage{hhline}
\usepackage{multirow}
\usepackage{multicol}
\usepackage{tikz-cd}
\usepackage{color}
\usepackage[super]{nth}
\usepackage{graphicx}
\usepackage{graphics}
\usepackage{listings}
\usepackage{import}
\usepackage{stmaryrd}
\usepackage{galois}
\usepackage{array}
\usepackage{pdflscape}
\usepackage{algorithmic}
\usepackage{galois}
\usepackage[ruled, linesnumbered]{algorithm2e}
\usepackage{hyperref}
\usepackage{capt-of}
\usepackage{makecell}
\usepackage{framed}
\usepackage{float}

\usepackage{graphicx}
\usepackage{subcaption}
\usepackage{caption}

\usepackage{pgf}
\usepackage{tikz, wrapfig,fancybox}
\usetikzlibrary{arrows,automata}
\usepackage[export]{adjustbox}



\makeatletter
\newenvironment{CenteredBox}{%
	\begin{Sbox}}{
	\end{Sbox}\centerline{\parbox{\wd\@Sbox}{\TheSbox}}}
\makeatother

\SetKwInput{KwData}{Input}
\SetKwInput{KwResult}{Output}

\lstdefinelanguage{JS}
{ keywords={int, main, while, this, return, if, else},
	otherkeywords={},
	basicstyle=\scriptsize\selectfont\ttfamily,
	keywordstyle=\bfseries\color{blue},
	sensitive=false,
	commentstyle=\color{purple!40!black},
	showspaces=false,
	tabsize=2,
	literate= {~}{\texttt{\phantom{m}}}1 {`}{$\texttt{\%}$}1 {?}{$\texttt{\$}$}1, 
	showstringspaces=false,emph={3}{\tiny }
	showtabs=true,
	morecomment=[l]{//},
	morecomment=[s]{/*}{*/},
	morestring=[b],
	breaklines=true,
	breakindent=12pt
}

\definecolor{dkred}{rgb}{.6,0,0}
\definecolor{dkgreen}{rgb}{0,.5,0}
\definecolor{dkblue}{rgb}{0,0,.6}
\definecolor{dkyellow}{cmyk}{0,0,.8,.3}
\definecolor{lightgray}{rgb}{.95,.95,.95}
\definecolor{darkgray}{rgb}{.3,.3,.3}
\definecolor{darkblue}{rgb}{0,0,.20}
\definecolor{purple}{rgb}{0.65, 0.12, 0.82}

\newif\ifdraft\draftfalse

\newcommand{\green}[2]{{\color{dkgreen}(#1: #2)}}
\newcommand{\xremark}[2]{{\color{dkred}(#1: #2)}}
\newcommand{\blu}[2]{{\color{purple}(#1: #2)}}
\ifdraft
\newcommand{\va}[1]{\xremark{VA}{#1}}
\newcommand{\ez}[1]{\blu{EZ}{#1}}
\newcommand{\im}[1]{\green{IM}{#1}}
\else
\newcommand{\va}[1]{}
\newcommand{\im}[1]{}
\newcommand{\ez}[1]{}
\fi

\newcommand{\defn}{\mathrel{\overset{\text{\rotatebox{-90}{$\triangleleft$}}}{=}}}
\newcommand{\lfp}{\mathrm{lfp}}

\newcommand{\Nset}{\mathbb{N}}
\newcommand{\Zset}{\mathbb{Z}}

\newcommand*{\fund}[3]{\mathord{#1}\colon#2\rightarrow#3}
\newcommand{\st}{\mathrel{.}}
\newcommand{\itc}{\mathrel{:}}

\newcommand*{\cond}[3]{({#1} \mathop{\mathrel{?}} {#2} \mathop{\mathrel{:}} {#3}) }

\newcommand{\ascdsc}[2]{{#1}\ensuremath{\uparrow\!\downarrow}{#2}}

\newcommand{\wid}{\nabla}
\newcommand{\nar}{\mathrm{\Delta}}
\newcommand{\ascdom}{\mathbb{A}}
\newcommand{\descdom}{\mathbb{D}}
\newcommand{\ascdesc}[2]{{\text{\ascdsc{$#1$}{$#2$}}}}
\newcommand{\ascdescdom}{\ascdesc{\ascdom}{\descdom}}

\newcommand{\alphaD}{\alpha_{\descdom}}
\newcommand{\gammaD}{\gamma_{\descdom}}
\newcommand{\alphaAD}{\alpha_{\ascdesc{\ascdom}{\descdom}}}
\newcommand{\gammaAD}{\gamma_{\ascdesc{\ascdom}{\descdom}}}

\newcommand{\WP}{\mathit{WP}}

\newcommand{\pardom}{\mathsf{Par}}
\newcommand{\lubpar}{\sqcup}
\newcommand{\glbpar}{\sqcap}
\newcommand{\botpar}{\bot}
\newcommand{\toppar}{\top}
\newcommand{\leqpar}{\sqsubseteq}
\newcommand{\pluspar}{\mathrel{+_\pardom}}

\newcommand{\intvdom}{\mathsf{Itv}}

\newcommand{\plusintv}{\mathrel{+_\intvdom}}

\newcommand*{\fn}{\mathrm{fn}}
\newcommand{\fndom}[1]{\mathsf{Set}_\fn({#1})}
\newcommand*{\wpfn}{\mathop{\wp_\fn}\nolimits}
\newcommand{\botfn}{\bot_\fn}

\newcommand{\lubfn}{\sqcup_\fn}
\newcommand{\leqfn}{\sqsubseteq_\fn}

\newcommand{\boxdom}{\mathsf{Box}}
\newcommand{\octdom}{\mathsf{Oct}}
\newcommand{\poldom}{\mathsf{Pol}}
\newcommand{\isetdom}{\mathsf{ISet}}

\newcommand{\psetdom}{\mathsf{PSet}}

\begin{document}

\title{Decoupling the ascending and descending phases in Abstract Interpretation}
\author{Vincenzo Arceri\inst{1} \and Isabella Mastroeni\inst{2} and Enea Zaffanella\inst{1}}
\authorrunning{V. Arceri, I. Mastroeni, E. Zaffanella}
\institute{
           University of Parma, Italy \\
           \email{\{vincenzo.arceri,enea.zaffanella\}@unipr.it}
\and
           University of Verona, Italy \\
           \email{isabella.mastroeni@univr.it}
}

\maketitle

\begin{abstract}
	Abstract Interpretation approximates the semantics of a program
	by mimicking its concrete fixpoint computation
	on an abstract domain $\ascdom$.
	The abstract (post-) fixpoint computation is classically divided
	into two phases:
	the \emph{ascending} phase, using widenings as extrapolation operators
	to enforce termination, is followed by a \emph{descending} phase,
	using narrowings as interpolation operators, so as to mitigate
	the effect of the precision losses introduced by widenings.
	In this paper we propose a simple variation of this classical approach
	where, to more effectively recover precision, we \emph{decouple}
	the two phases: in particular, before starting the descending phase,
	we replace the domain $\ascdom$ with a more precise abstract
	domain $\descdom$.
	The correctness of the approach is justified by casting it
	as an instance of the A$^2$I framework.
	After demonstrating the new technique on a simple example,
	we summarize the results of a preliminary experimental evaluation,
	showing that it is able to obtain significant precision improvements
	for several choices of the domains $\ascdom$ and $\descdom$.
\end{abstract}


\section{Introduction}\label{sect:intro}
Abstract interpretation \cite{CC77} is a framework for designing
approximate semantics, with the aim of gathering information about
programs in order to provide conservative/sound answers to questions
about their run-time behaviors. In other words, the purpose of
abstract interpretation is to formally design automatic program
analyses by approximating program semantics for statically determining
dynamic properties.
The design of static analyzers consists in automatizing the computation
of such approximations, and in this case the answer can only be
partial or imprecise, due to the undecidability of program
termination.
Abstract/approximated semantics are computed by
mimicking the monotonic (ascending) concrete semantics computation,
obtained by Kleene iteration reaching fixpoint. Unfortunately, it is
well known that Kleene fixpoint computation may not terminate. In the
static analysis framework this issue has been tackled by introducing
fixpoint accelerators, namely new operators (called widenings) built
on the computational abstract domain, allowing to accelerate the
fixpoint computation at the price of potentially reaching a
post-fixpoint, namely at the price of losing precision in the answer.
For this reason, it is common in static analysis to design another
operator (called narrowing) performing a descending path in order to
try to recover some precision by refining the reached post-fixpoint.

The precision of the result depends both on the ability of the
widening operator to \emph{guess} a limit of the increasing sequence, and
on the information gathered during the decreasing
phase. Intuitively, the increasing sequence extrapolates the
behavior of the program from the first steps of its execution, while
the decreasing sequence gathers information about the end of the
execution of the program \cite{BoutonnetH18}. Moreover, a naive
application of the classical approach may lead to an inadequate
analysis, which is too expensive or too imprecise, meaning that there
is a strong need for mechanisms that can effectively tune the
precision/efficiency tradeoff.  In order to improve this ration, we
could either improve efficiency (usually to the detriment of
precision) by choosing a simpler (less precise) domain or by changing
the fixpoint construction (e.g., replacing precise abstract operators
with cheaper over-approximations), or improve precision (usually to
the detriment of efficiency) by choosing a more precise/costly domain
or again by changing the fixpoint construction (clearly in the
opposite direction, see techniques discussed in Section~\ref{sect:rel}).

In this work, we propose to combine these improvement approaches by
choosing to use different domains (with different precision degrees)
depending on the analysis phase: we use a potentially less
precise domain in the fixpoint computation exploiting a widening
operator for reaching a post-fixpoint in the ascending phase, and
therefore potentially sensitively losing precision, and we use a more
precise domain in the descending (narrowing) phase for trying to
improve the gain of precision of such phase.
The idea is rather simple but, to the best of our knowledge, it was
never proposed before; also, since it is orthogonal with respect to
similar approaches, it can be used in combination with them (rather
than as an alternative to them).  The intuition beyond the gain of
precision without a relevant loss of efficiency is based on the idea
that in the descending phase we do not need to use the more expensive
operations. Such intuition is supported by our initial experimental
evaluation, showing that the proposed approach is surely promising,
being able to improve precision in a significant number of cases.

\paragraph*{Paper structure.}
Section~\ref{sect:bg} gives basics in order theory, Abstract
Interpretation, and the classical approach for static analysis by
Abstract Interpretation. Section~\ref{sect:asc-desc} presents our
proposal for decoupling the ascending and descending phases with two
different abstract domains. Section~\ref{sect:expeval} reports a
preliminary experimental evaluation of our
approach. Section~\ref{sect:rel} discusses most related
works. Section~\ref{sect:concl} concludes.

\section{Background}\label{sect:bg}
\paragraph*{Order theory.}
We denote by $\wp(S)$ the powerset of a set $S$.
A poset $\langle L, \sqsubseteq_L \rangle$ is a set $L$ equipped
with a partial order $\mathord{\sqsubseteq_L} \in \wp(L \times L)$,
i.e., a reflexive, transitive and anti-symmetric binary relation;
in the following we will omit subscripts when clear from context.
A poset is a join semi-lattice if, for each $l_1, l_2 \in L$,
the lub (least upper bound) $l_1 \mathrel{\sqcup} l_2$ belongs to $L$;
similarly, it is a meet semi-lattice if the glb (greatest lower bound)
$l_1 \mathrel{\sqcap} l_2$ belongs to $L$; when both properties hold,
we have a lattice $\langle L, \sqsubseteq, \sqcup, \sqcap \rangle$.
A lattice is complete if $\forall X \subseteq L$,
$\bigsqcup X$ and $\bigsqcap X$ belong to $L$;
a complete lattice with bottom element $\bot$ and top element $\top$
is denoted $\langle L, \sqsubseteq, \sqcup, \sqcap, \bot, \top \rangle$.
A poset $\langle L, \sqsubseteq \rangle$
satisfies the ascending chain condition (ACC) iff each infinite sequence
$l_0 \sqsubseteq l_1 \sqsubseteq \dots \sqsubseteq l_i \sqsubseteq \dots$
of elements of $L$ is not strictly increasing,
i.e., $\exists k \geq 0, \forall j \geq k \itc l_k = l_j$.
Dually the poset satisfies the descending chain condition (DCC)
iff each infinite sequence
$l_0 \sqsupseteq l_1 \sqsupseteq \dots \sqsupseteq l_i \sqsupseteq \dots$
of elements of $L$ is not strictly decreasing, that is
$\exists k \geq 0, \forall j \geq k \itc l_k = l_j.$

A function $\fund{f}{L}{L}$ on poset $\langle L, \sqsubseteq \rangle$
is monotone if, for all $l_1, l_2 \in L$,
$l_1 \sqsubseteq l_2$ implies $f(l_1) \sqsubseteq f(l_2)$.
We denote $\mathrm{post}(f)$ the set of post-fixpoints of $f$,
i.e., those elements $x \in L$ satisfying $f(x) \sqsupseteq x$;
similarly, $\mathrm{pre}(f)$ is the set of pre-fixpoints of $f$,
satisfying $f(x) \sqsubseteq x$; the set of fixpoints of $f$,
satisfying $f(x) = x$, is thus
$\mathrm{fix}(f) = \mathrm{pre}(f) \cap \mathrm{post}(f)$.
Given a function $\fund{f}{L}{L}$ we recursively define
the iterates/iterations of $f$ from $x \in L$
as $f^0(x) = x$ and $f^{i+1}(x) = f(f^i(x))$.
The Kleene fixpoint theorem says that
a continuous function $\fund{f}{L}{L}$ on a complete lattice
$\langle L, \sqsubseteq, \sqcup, \sqcap, \bot, \top \rangle$
has a least fixpoint $\lfp(f) \in L$,
which can be obtained as the lub of the increasing sequence
$f^0(\bot) \sqsubseteq f^1(\bot) \sqsubseteq \dots
   \sqsubseteq f^i(\bot) \sqsubseteq \dots$~\cite{CC79}.

\paragraph*{Abstract Interpretation (AI).}
Abstract Interpretation \cite{CC77,CC79}
is a theory to soundly approximate program semantics,
focusing on some run-time property of interest.
In the classical setting, the concrete and the abstract semantics are
defined over two complete lattices,
respectively called the concrete domain $C$ and the abstract domain $A$.
A pair of monotone functions $\alpha: C \rightarrow A$ and
$\gamma: A \rightarrow C$ forms a \emph{Galois Connection} (GC)
if
\(
\forall c \in C, \forall a \in A
\itc
\alpha(c) \sqsubseteq_A a \Leftrightarrow c \sqsubseteq_C \gamma(a)
\).
If $C$ and $A$ are related by a GC, denoted $C \galois{\alpha}{\gamma} A$,
then an abstract function $\fund{f_A}{A}{A}$
is a correct approximation of a concrete function $\fund{f_C}{C}{C}$
if and only if
$\forall c \in C \itc \alpha(f_C(c)) \sqsubseteq_A f_A(\alpha(c))$ or equivalently $\forall a \in A \itc
	f_C(\gamma(a))
	\sqsubseteq_C \gamma(f_A(a))$;
the \emph{best correct approximation} of $f_C$ is
$f_A^\sharp = (\alpha \circ f_C \circ \gamma)$.

\paragraph{Static program analysis via abstract interpretation.}
It is possible to represent a
program of interest as a control-flow graph (CFG for short). A CFG is a graph $\langle N, E \rangle$
such that $N = \{n_1, n_2, \dots, n_m\}$ is a finite set of nodes corresponding to the control points of
the program, and $E \subseteq N \times N$ is a finite set of edges. It is possible to compute the CFG
associated with a certain program with standard techniques~\cite{popa}.

Let us denote by $A$ the abstract domain approximating the concrete domain $C$, used to analyze
programs of interest. With each node $n \in N$ is associated a function transformer $f_n : A^m
	\rightarrow A$ capturing the effects of the node $n$, i.e., the abstract semantics. Analyzing a given
CFG $C = \langle N, E\rangle$, where $N = \{n_1, n_2, \dots, n_m\}$ means to resolve the 
following system of equations%
\footnote{In general, the least fixpoint on the concrete domain $C$
is not finitely computable. Hence, the idea is to compute an abstract fixpoint,
over an abstract domain $A$, that correctly approximates the concrete one.}
%
\[
  F = \bigl\{\,
        x_i = f_i(x_1, x_2, \dots, x_m)
      \bigm|
        i = 1, 2, \dots, m
      \,\bigr\}
\]
The goal of AI-based static analysis,
using an abstract domain $A$, is to compute the least solution
of the equation set $F$ as the limit of a Kleene iteration on $A$,
i.e.,
$x^{\lfp(F)} \defn (x_1^{\lfp(F)}, \dots, x_m^{\lfp(F)})$
starting from the bottom elements of $A$,
i.e., $\forall i \in [1, m].\: x_i = \bot$.

\begin{example}
\label{ex:parity-example}
\begin{figure}[t]
\begin{subfigure}[b]{0.5\textwidth}
\begin{CenteredBox}
\begin{lstlisting}
int main() {
	int x = 0;
	while (x < 100)
	if (x < 50)
		x = x + 2;
	else
		x = x + 10;
}
\end{lstlisting}
\end{CenteredBox}
\caption{}
\label{fig:cfg-example-a}
\end{subfigure}
~
\begin{subfigure}[b]{0.5\textwidth}
\centering
\includegraphics[scale=0.3,bb=0 0 540 406]{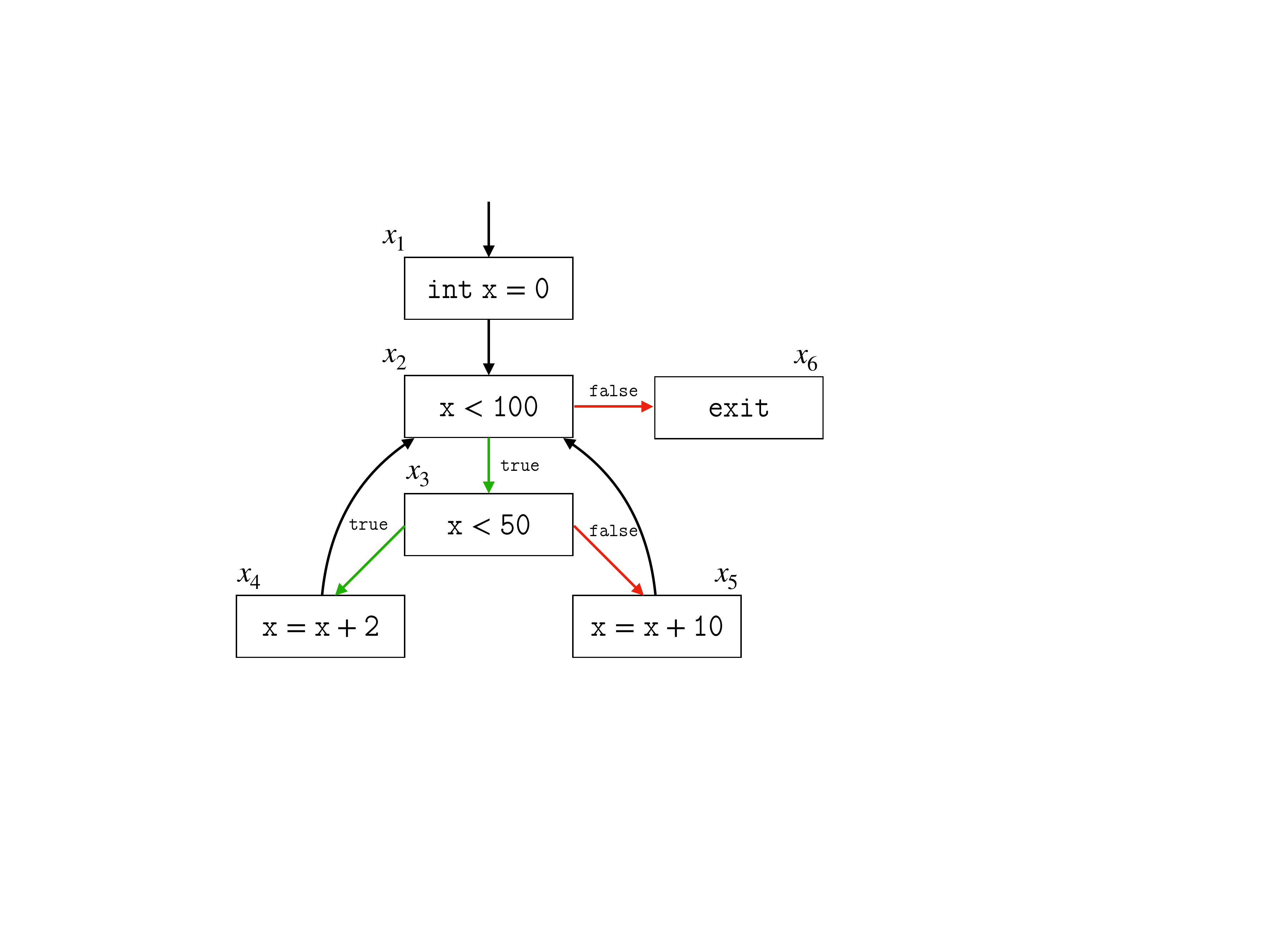}
\caption{}
\label{fig:cfg-example-b}
\end{subfigure}
~
\begin{subfigure}[b]{1\textwidth}
\begin{align*}
x_1 &= \top \qquad x_3 = x_2 \qquad x_4 = x_3
            \qquad x_5 = x_3 \qquad x_6 = x_2 \\
x_2 &= \mathsf{Even} \sqcup (x_4 \pluspar \mathsf{Even})
                     \sqcup (x_5 \pluspar \mathsf{Even})
\end{align*}
\caption{}
\label{fig:cfg-example-c}
\end{subfigure}
\caption{(a) C function example, (b) associated CFG,
(c) associated system of equations with the $\pardom$ abstract domain.}
\label{fig:cfg-example}
\end{figure}
	Consider the C function of Figure~\ref{fig:cfg-example-a}
	and the corresponding CFG, shown in Figure~\ref{fig:cfg-example-b}.
	We intuitively describe the analysis of this program using
	the abstract domain $\pardom$~\cite[Example 10.1.0.3]{CC79},
	tracking the parity of numerical variables:
	\[
		\pardom = \bigl\langle
		\{\botpar, \toppar, \mathsf{Even}, \mathsf{Odd}\},
		\leqpar, \lubpar, \glbpar, \botpar, \toppar
		\bigr\rangle,
	\]
	where the partial order is defined by
	$\botpar \leqpar x \leqpar \toppar$, for each $x \in \pardom$.
	The system of equations is reported in Figure~\ref{fig:cfg-example-c};
	Note that the equation defining $x_i$ is intuitively describing
	the values that are possibly \emph{entering} the corresponding node
	of the CFG (also labeled $x_i$ for convenience);
	for instance, the right hand side of the equation defining $x_2$
	computes the lub of the abstract values \emph{exiting}
	from nodes $x_1$, $x_4$ and $x_5$, respectively.
	For space reasons, we leave to intuition the abstract functions modeling
	the semantics of each CFG node
	(e.g., function $\fund{\pluspar}{\pardom \times\pardom}{\pardom}$
	modeling addition on the $\pardom$ domain).
	The least solution for the system is $x_1 = \toppar$ and
	$x_i = \mathsf{Even}$ for $i = 2$, \dots, $6$.
	\qed
\end{example}

The ascending sequence over the system of equations $F$
may fail to (finitely) converge for abstract domains
that do not satisfy the ACC.
A converge guarantee can be provided by widening operators, which
over-approximate the least fixpoint solution $x^{\lfp(F)}$
by effectively computing a post-fixpoint of $F$.
A widening $\fund{\wid}{A \times A}{A}$ is an operator such that:
\begin{itemize}
  \item for each $a_1, a_2 \in A$,
        $a_1 \sqsubseteq a_1 \mathrel{\wid} a_2$ and
        $a_2 \sqsubseteq a_1 \mathrel{\wid} a_2$;
  \item for all ascending sequences
        $a_0 \sqsubseteq \dots \sqsubseteq a_{i+1} \sqsubseteq \dots$,
        the ascending sequence
        $x_0 \sqsubseteq \dots \sqsubseteq x_{i+1} \sqsubseteq \dots$
	defined by $x_0 = a_0$ and $x_{i + 1} = x_i \mathrel{\wid} a_{i + 1}$
	is not strictly increasing.
\end{itemize}
In principle, widening can be applied to all equations of the
system $F$, which however would lead to a gross over-approximation;
following~\cite{Bourdoncle93}, it is sufficient that the widening is
applied on one node in each cycle of the CFG;
for instance, in Figure~\ref{fig:cfg-example-b}
we can use $x_3$ as the one and only widening point.
We denote $\WP \subseteq N$ the set of \emph{widening points},
i.e., the nodes of the CFG where widening is applied, leading
to the system of equations $F^\wid$:
\begin{equation}\label{eq:eq-wid}
  \begin{cases}
   x_i = x_i \mathrel{\wid} f_i(x_1, x_2, \dots, x_m),
     & \text{if $i \in \WP$;} \\
   x_i = x_i \mathrel{\sqcup} f_i(x_1, x_2, \dots, x_m),
     & \text{otherwise.}
   \end{cases}
\end{equation}

\begin{figure}[t]
  \centering
  \includegraphics[scale=0.4,bb=0 0 374 570]{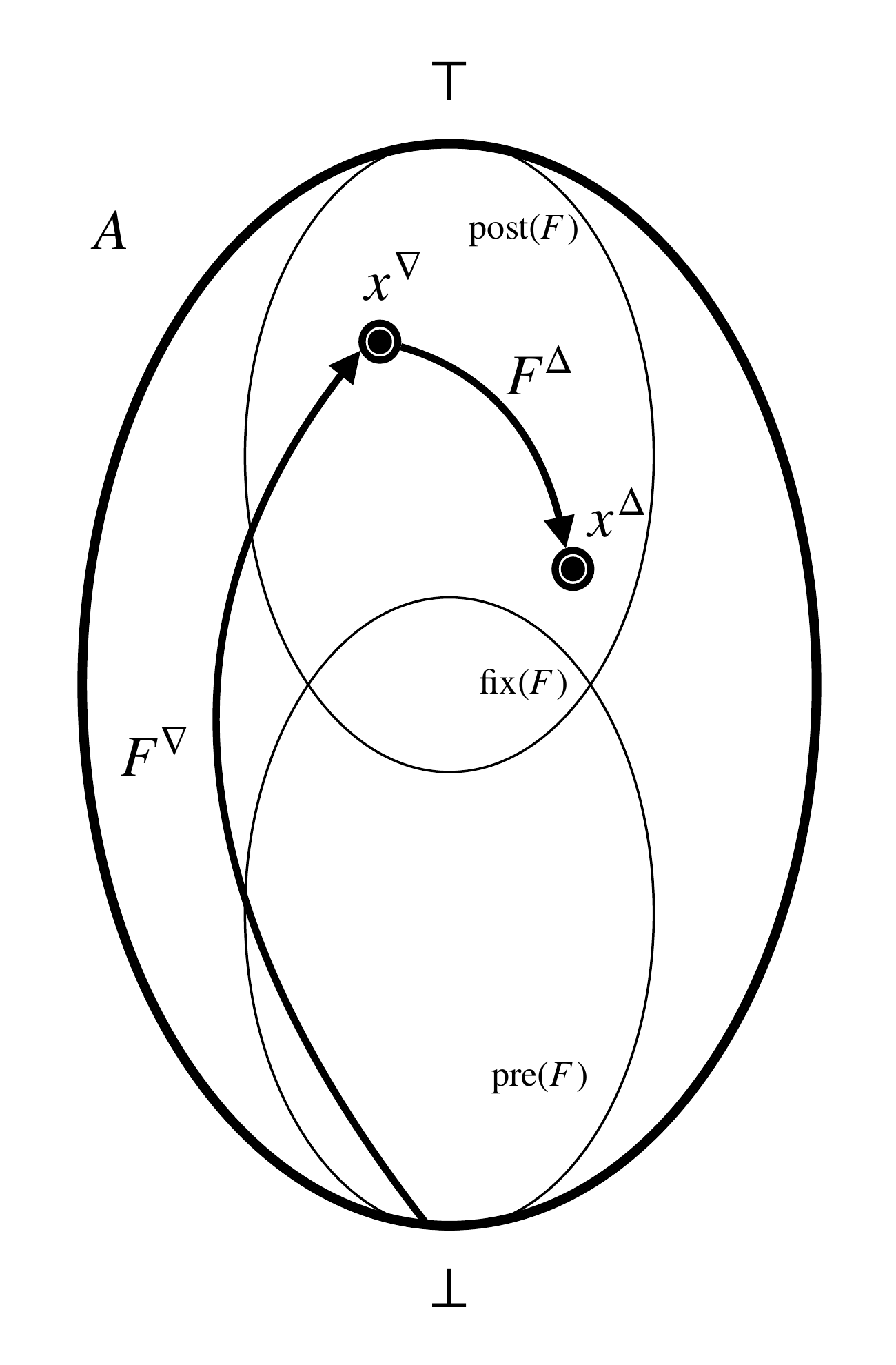}
  \caption{The ascending and descending phases over abstract domain $A$.}
  \label{fig:asc-desc}
\end{figure}

In order to mitigate the loss of precision introduced by widenings,
the \emph{ascending phase} computing the post-fixpoint $x^\wid$ of $F$
can be followed by another Kleene iteration on the system $F$,
starting from $x^\wid$ and descending towards
a fixpoint of $F$ (not necessarily the least one).
If the abstract domain $A$ does not satisfy the DCC, this descending sequence
may fail to converge; a convergence guarantee can be obtained
by using a narrowing operator $\fund{\nar}{A \times A}{A}$, satisfying:
\begin{itemize}
  \item for each $a_1, a_2 \in A$,
	$a_1 \sqsupseteq a_1 \mathrel{\nar} a_2
             \sqsupseteq a_1 \mathrel{\sqcap} a_2$;
  \item for all descending sequences
        $a_0 \sqsupseteq \dots \sqsupseteq a_{i+1} \sqsupseteq \dots$,
	the descending sequence
        $x_0 \sqsupseteq \dots \sqsupseteq x_{i+1} \sqsupseteq \dots$
        defined by $x_0 = a_0$ and $x_{i + 1} = x_i \mathrel{\nar} a_{i + 1}$
        is not strictly decreasing.
\end{itemize}
As before, the application of narrowings can be limited to $\WP$,
leading to the system of equations $F^\nar$ used during the
\emph{descending phase}:
\begin{equation}\label{eq:eq-narr}
  \begin{cases}
   x_i = x_i \mathrel{\nar} f_i(x_1, x_2, \dots, x_m),
     & \text{if $i \in \WP$;} \\
   x_i = x_i \mathrel{\sqcap} f_i(x_1, x_2, \dots, x_m),
     & \text{otherwise.}
   \end{cases}
\end{equation}
In general, the descending sequence with narrowing
will compute a post-fixpoint $x^\nar$ of $F$
(not necessarily a fixpoint), satisfying $x^\nar \sqsubseteq x^\wid$.
A graphical representation of the ascending and descending phases over the
abstract domain $A$ is reported in Figure~\ref{fig:asc-desc}.
Note that a ``glb-based'' narrowing operator can be easily defined
by computing the domain glb and forcing the descending sequence
to stop as soon as reaching a fixed, finite number $k \in \Nset$
of iterations. For this reason, several abstract domains
do not implement a proper narrowing operator.

\begin{example}
\label{ex:interval-example}
\begin{figure}[t]
\begin{subfigure}[b]{1\textwidth}
  \begin{align*}
  &x_1 = \top
    & \qquad & x_4 = x_3 \sqcap [-\infty, 49] \\
  &x_2 = [0,0] \sqcup (x_4 \plusintv [2,2]) \sqcup (x_5 \plusintv [10,10])
    & \qquad & x_5 = x_3 \sqcap [50, +\infty]  \\
  &x_3 = x_2 \sqcap [-\infty, 99]
    & \qquad & x_6 = x_2 \sqcap [100, +\infty]
  \end{align*}
\caption{}
\label{fig:intv-analysis-ex-a}
\end{subfigure}
~
\begin{subfigure}[b]{1\textwidth}
\centering
\begin{tabular}{|c|c|c|c||c|c|}
\multicolumn{2}{c|}{}
   & \multicolumn{2}{c||}{ascending phase iter}
   & \multicolumn{2}{c|}{descending phase iter} \\
\hline
$N$ & $\WP$ & 1st & 2nd ($= x^\wid_\intvdom$)
            & 1st & 2nd ($= x^\nar_\intvdom$) \\
\hline
$x_1$ & & $\top$ & $\top$
        & $\top$ & $\top$ \\
$x_2$ & & $[0,0]$              & $[0, +\infty]$
        & $[0, +\infty]$       & $[0, 109]$ \\
$x_3$ & $\checkmark$
      & $[0,0]$         & $[0, +\infty]$
      & $[0,99]$             & $[0,99]$ \\
$x_4$ & & $[0, 0]$             & $[0,49]$
        & $[0, 49]$            & $[0,49]$ \\
$x_5$ & & $\bot$               & $[50, +\infty]$
        & $[50, 99]$           & $[50, 99]$ \\
$x_6$ & & $\bot$               & $[100, +\infty]$
        & $[100, +\infty]$     & $[100, 109]$ \\
\hline
\end{tabular}
\caption{}
\label{fig:intv-analysis-ex-b}
\end{subfigure}
\caption{(a) Equations for CFG in Figure~\ref{fig:cfg-example-b}
         using $\intvdom$, (b) interval results.}
\label{fig:intv-analysis-ex}
\end{figure}

The domain of integral intervals~\cite{CC92b}
(or 1-dimension integral boxes)
is an example of complete lattice satisfying neither the ACC nor the DCC:
\[
  \intvdom
    = \{ \bot, \top \}
    \cup \bigl\{ [\ell, u] \bigm| \ell, u \in \Zset, \ell \leq u \bigr\}
    \cup \bigl\{ [-\infty, u] \bigm| u \in \Zset \bigr\}
    \cup \bigl\{ [\ell, +\infty] \bigm| \ell \in \Zset \bigr\},
\]
where $\bot$ is the bottom element (denoting the empty interval),
$\top = [-\infty, +\infty]$ is the top element (denoting $\Zset$)
and the partial order, lub and glb operators consistently model
the usual containment relation.
The interval widening operator~\cite{CC92b}
$\fund{\wid}{\intvdom \times \intvdom}{\intvdom}$ is defined,
for each $x \in \intvdom$, by
$\bot \mathrel{\wid} x = x \mathrel{\wid} \bot = x$ and
\begin{equation*}
  [\ell_0, u_0] \mathrel{\wid} [\ell_1, u_1]
    = [ \cond{\ell_1 < \ell_0}{-\infty}{\ell_0},
        \cond{u_0 < u_1}{+\infty}{u_0} ].
\end{equation*}
Similarly, the interval narrowing operator~\cite{CC92b}
$\fund{\nar}{\intvdom \times \intvdom}{\intvdom}$ is defined,
for each $x \in \intvdom$, by
$\bot \mathrel{\nar} x = x \mathrel{\nar} \bot = \bot$ and
\begin{equation*}
  [\ell_0, u_0] \mathrel{\nar} [\ell_1, u_1]
    = [ \cond{\ell_0 = -\infty}{\ell_1}{\ell_0},
        \cond{u_0 = +\infty}{u_1}{u_0} ].
\end{equation*}

Considering again the C function in Figure~\ref{fig:cfg-example-a},
the corresponding system of equations for the domain $\intvdom$
is shown in Figure~\ref{fig:intv-analysis-ex-a}
(where $\fund{\plusintv}{\intvdom \times \intvdom}{\intvdom}$
models addition on the $\intvdom$ domain).
The computation of the ascending and descending sequences
is shown in Figure~\ref{fig:intv-analysis-ex-b},
where in the 2nd column we have highlighted the only widening point $x_3$;
in particular, the 4th and 6th columns show the post-fixpoint and
the fixpoint obtained at the end of the ascending and the descending phases,
respectively.
	\qed
\end{example}

\paragraph*{Powerset domains.}
Many abstract domains (e.g., numerical domains whose elements are convex sets)
are unable to precisely describe disjunctive information, thereby incurring
significant precision losses whenever the abstract semantic construction
needs to merge different control flow paths.
To avoid these losses, it is possible to lift the domain
using a disjunctive domain refinement operator~\cite{CC79}.
In the following we will consider the finite powerset~\cite{BagnaraHZ06}
of an abstract domain $A$, which is the join-semilattice
$\fndom{A} = \langle \wpfn(A), \leqfn, \lubfn, \botfn \rangle$,
where:
\begin{itemize}
\item
the carrier $\wpfn(A)$ is the set of the \emph{finite}
and \emph{non-redundant} subsets of $A$
(an element $a_1 \in A$ is redundant in $S \subseteq A$ iff
$a_1 = \bot_A$ or $\exists a_2 \in S \st a_1 \sqsubset_A a_2$);
\item
the partial order $S_1 \leqfn S_2$ is defined by
$\forall a_1 \in S_1, \exists a_2 \in S_2 \st a_1 \sqsubseteq_A a_2$;
\item
the (binary) least upper bound $S_1 \lubfn S_2$
is computed by removing the redundant elements from
the set union $S_1 \cup S_2$;
\item
the bottom element is $\botfn = \emptyset$.
\end{itemize}
For space reasons we omit a more thorough discussion of powerset
domains (e.g., the lifting of the abstract semantic operators
defined on $A$), referring the interested reader to~\cite{BagnaraHZ06,CC79}.

\section{Decoupling the ascending and descending phases}\label{sect:asc-desc}
In the previous section we have recalled the classical approach used
in static analysis based on abstract interpretation, which can be
summarized as follows:
(a) fix an abstract domain $A$ such that $C \galois{\alpha}{\gamma} A$
and a corresponding, correct system of abstract equations $F_A$;
(b) approximate the concrete semantics by computing a post-fixpoint
of $F_A$ in the ascending phase (with widening);
(c) improve the result in the descending phase (with narrowing).
What is worth noting is that the two phases (b) and (c)
are computed on the same domain $A$.

Moving from the observation that the only goal of the descending phase
is to improve precision, we propose to \emph{decouple} it from the
ascending phase: that is, we compute the descending sequence
on a different, more precise abstract domain, so as to increase
the chances of a significant precision improvement.
Clearly, the adoption of a more precise domain likely incurs
some penalty in terms of the efficiency of the analysis; however,
since in our proposal this domain is only used in the descending phase,
it should be simpler to achieve a good tradeoff between precision
and efficiency, because the descending phase can be stopped after any
number of iterations and still provide a correct result.

In the following we will denote $\ascdom$ and $\descdom$ the
abstract domains used in the ascending and descending phases, respectively,
and use the notation $\ascdescdom$ to refer to this decoupled approach.
The correctness/precision relation between the concrete domain
and the two abstract domains is formalized by requiring
$C \galois{\alphaD}{\gammaD} \descdom \galois{\alphaAD}{\gammaAD} \ascdom$;
we also require that the concretization function $\gammaAD$
is effectively computable.

\begin{figure}[t]
\centering
\includegraphics[scale=0.4,bb=0 0 844 583]{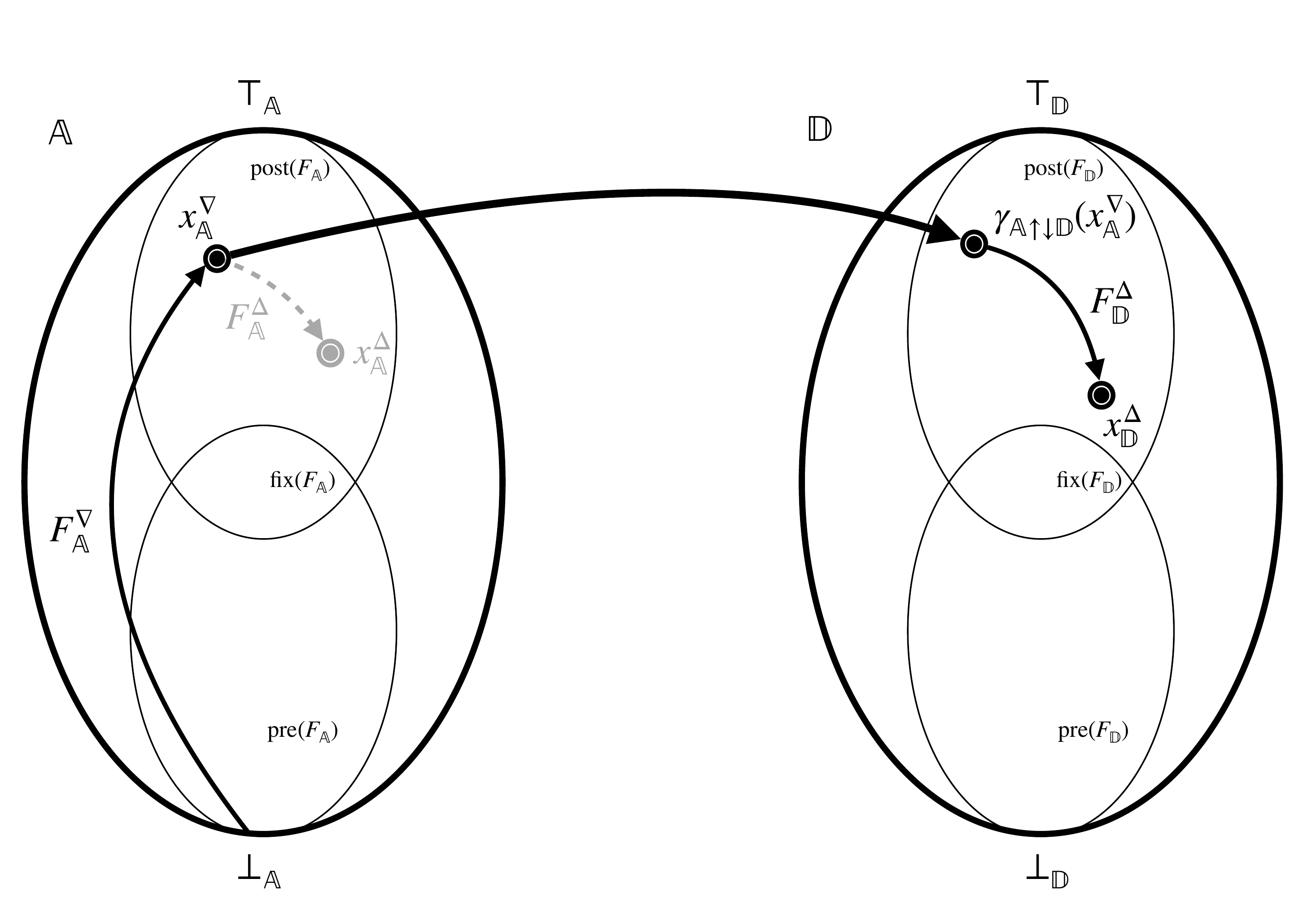}
\caption{The ascending and descending phases over $\ascdom$ and
         $\descdom$, respectively.}
\label{fig:asc-desc-split}
\end{figure}

Our decoupled approach is graphically represented in
Figure~\ref{fig:asc-desc-split}.
The (concrete) system of equations $F_C$
is correctly approximated on domain $\descdom$ by
the (abstract) system of equations $F_\descdom$,
which is further approximated on domain $\ascdom$ by
the system of equations $F_\ascdom$.
We first compute a post-fixpoint $x^\wid_{\ascdom} \in \ascdom$
using the system of equations $F_{\ascdom}^\wid$ (with widening);
instead of descending on the same abstract domain,
as done in Section~\ref{sect:bg},
we transfer the post-fixpoint $x^\wid_{\ascdom}$
to the more precise domain $\descdom$,
using the concretization function $\gammaAD$ (which is computable);
hence, the descending phase will use the system of equations
$F^\nar_{\descdom}$ (with narrowing) on domain $\descdom$,
starting from $\gammaAD(x^\wid_{\ascdom})$ and
obtaining an improved post-fixpoint $x^\nar_{\descdom} \in \descdom$.

The next lemma states that the post-fixpoint $x^\wid_{\ascdom} \in \ascdom$
corresponds to a post-fixpoint for $F_{\descdom}$,
necessary for starting the descending phase on $\descdom$.

\begin{lemma}
	\label{lem:postfixpoint-transfer-alt}
	Consider
	$\descdom \galois{\alpha_{\ascdescdom}}{\gamma_{\ascdescdom}} \ascdom$
	and let
	$\fund{F_{\ascdom}}{\ascdom}{\ascdom}$
	be a correct approximation of
	$\fund{F_{\descdom}}{\descdom}{\descdom}$. Then,
	\[
	x^\wid_{\ascdom} \in \mathrm{post}(F_{\ascdom})
	\implies
	\gamma_{\ascdescdom}(x^\wid_{\ascdom}) \in \mathrm{post}(F_{\descdom}).
	\]
\end{lemma}

It would be desirable to prove that the final result $x^\nar_\descdom$
obtained when using the decoupled approach $\ascdescdom$
systematically improves on
the final result $x^\nar_\ascdom$
obtained by the classical approach, i.e.,
$x^\nar_\descdom \mathrel{\sqsubseteq_\descdom} \gammaAD(x^\nar_\ascdom)$.
However, in general this property does not hold, due to the use of
different, unrelated, possibly non-monotonic narrowing operators
on the domains $\ascdom$ and $\descdom$.
We can prove the desired result provided we force both domains to use
the glb-based narrowing (with the same threshold value).

\begin{proposition}\label{prop:prec-alt}
Consider $\descdom \galois{\alphaAD}{\gammaAD} \ascdom$
and let
$\fund{F_{\ascdom}}{\ascdom}{\ascdom}$
be a correct approximation of
$\fund{F_{\descdom}}{\descdom}{\descdom}$;
let also $x^\wid_{\ascdom} \in \mathrm{post}(F_{\ascdom})$.
Then, for each $k \in \mathbb{N}$,
\[
  F_{\descdom}^k(\gammaAD(x^\wid_\ascdom))
    \mathrel{\sqsubseteq_\descdom}
      \gammaAD(F_{\ascdom}^k(x^\wid_\ascdom)).
\]
\end{proposition}

Note that Lemma~\ref{lem:postfixpoint-transfer-alt}
and Proposition~\ref{prop:prec-alt} are well-known results.
Intuitively, the correctness of the decoupled approach is easily
justified by viewing it as an instance of
the A$^2$I framework~\cite{CousotGR19}:
starting from a classical analysis using the more precise domain $\descdom$,
we further abstract part of its computation (the ascending phase),
approximating it on domain $\ascdom$.

\paragraph*{On the Galois Connection requirement.}
When formalizing our decoupled proposal, we have assumed that
all the considered domains (concrete, ascending and descending)
are related by GCs:
this corresponds to an ideal situation where
for each element $x$ of the more precise domain
(resp., each semantic transformer $f$)
we can identify the corresponding best correct approximation
on the less precise domain
$\alpha(x)$ (resp., $\alpha \circ f \circ \gamma$).
However, there are well-known abstract domains
(e.g., the domain convex polyhedra~\cite{CousotH78}
approximating sets of reals or
the deterministic finite-state automata domain~\cite{Arceri20}
approximating sets of strings)
that cannot be related to the concrete domain using a GC.
This is not a real concern because, as discussed at length in~\cite{CC92},
one can adopt a slightly weaker theoretical framework and still ensure
the correctness of the analysis.
As a matter of fact, in the experimental evaluation
we will implicitly relax the GC assumption.

\begin{example}
We now reconsider the C program in Figure~\ref{fig:cfg-example-a}
and show how the decoupled approach can be used to improve on
the results computed by the classical analysis on domain $\intvdom$
(see Example~\ref{ex:interval-example} and
Figure~\ref{fig:intv-analysis-ex-b}).
To this end, while keeping the domain $\intvdom$ for the ascending phase,
we will compute the descending phase on
the powerset domain $\isetdom = \fndom{\intvdom}$,
i.e., we will adopt the combination \ascdsc{$\intvdom$}{$\isetdom$}.

\begin{figure}[t]
\centering
\begin{tabular}{|c|c||c|c|c|}
\hhline{~---~}
\multicolumn{1}{c|}{}
  & transfer
  & \multicolumn{2}{c|}{descending phase}
  & \multicolumn{1}{c}{} \\
\hhline{----~}
$N$ & $\gamma_{\ascdsc{\intvdom}{\isetdom}}(x^\wid_{\intvdom})$
    & 1st iter & 2nd iter & \multicolumn{1}{c}{} \\
\hline
$x_1$ & $\{ \top_\intvdom \}$
      & $\{ \top_\intvdom \}$
      & $\{ \top_\intvdom \}$
      & \\
$x_2$ & $\{ [0,+\infty] \}$
      & $\{ [0,0], [2,51], [60,+\infty] \}$
      & $\{ [0,0], [2,2], [4,51], [60,61], [70,109] \}$
      & $\checkmark$ \\
$x_3$ & $\{ [0,+\infty] \}$
      & $\{ [0,0], [2,51], [60,99] \}$
      & $\{ [0,0], [2,2], [4,51], [60,61], [70,99] \}$
      & $\checkmark$ \\
$x_4$ & $\{ [0,49] \}$
      & $\{ [0,0], [2,49] \}$
      & $\{ [0,0], [2,2], [4,49] \}$
      & $\checkmark$ \\
$x_5$ & $\{ [50,99] \}$
      & $\{ [50,51], [60,99] \}$
      & $\{ [50,51], [60,61], [70,99] \}$
      & $\checkmark$ \\
$x_6$ & $\{ [100,+\infty] \}$
      & $\{ [100,+\infty] \}$
      & $\{ [100,109] \}$
      & \\
\hline
\end{tabular}
\caption{Computing the descending phase on \ascdsc{$\intvdom$}{$\isetdom$}.}
\label{fig:pintv-analysis-ex}
\end{figure}

Before starting the descending phase,
the post-fixpoint $x^\wid_\intvdom$
computed in the ascending phase using $\intvdom$
(see the 4th column of Figure~\ref{fig:intv-analysis-ex-b})
is transferred to $\isetdom$ using
$\fund{\gamma_{\ascdsc{\intvdom}{\isetdom}}}{\intvdom}{\isetdom}$,
obtaining the (singleton) sets of intervals shown in the 3rd column of
Figure~\ref{fig:pintv-analysis-ex}.
Then the descending phase on $\isetdom$ is started:
note that we use the glb-based narrowing,
with a threshold value $k = 2$ on the number of iterations;
the results computed by the two iterations are shown
in the 3rd and 4th columns of Figure~\ref{fig:pintv-analysis-ex}.

It is now possible to perform a precision comparison of the results
obtained on domain $\intvdom$ using the classical approach
(last column in Figure~\ref{fig:intv-analysis-ex-b})
with respect to the results obtained with the
\ascdsc{$\intvdom$}{$\isetdom$} combination
(4th column of Figure~\ref{fig:pintv-analysis-ex}):
for convenience, in the last column we show a checkmark ($\checkmark$)
on the CFG nodes where we actually obtain a precision improvement.
Note that the post-fixpoint computed on $\isetdom$ is not a fixpoint:
hence, the precision could be refined further by increasing
the threshold value $k \in \Nset$.
\qed
\end{example}

\section{Experimental evaluation}\label{sect:expeval}
In order to obtain a preliminary experimental evaluation
of the precision gains resulting from the proposed analysis technique,
we have modified the open source static analysis tool PAGAI~\cite{HenryMM12}
to allow for decoupling the ascending and descending iteration phases;
in particular, we have added program options to select a different abstract
domain for the descending phase, as well as to select a threshold value
for the number of descending iterations (this threshold is set to 3).%
\footnote{By design, PAGAI does not use proper narrowing operators
to enforce the termination of the decreasing sequence;
rather, it stops when the iteration count reaches the threshold value
(or earlier, if a fixpoint is detected).}
In our experiments we configured PAGAI to perform
a \emph{simple} static analysis: hence,
we disregard more sophisticated approaches, such as \emph{path focusing},
and we disabled those LLVM bitcode instrumentation passes
that heavily modify the CFG in order to potentially detect
overflows and other runtime errors.

PAGAI is interfaced with the Apron library~\cite{JeannetM09},
which provides several numeric domains,
among which boxes~\cite{CC77} ($\boxdom$),
octagons~\cite{Mine06} ($\octdom$) and
convex polyhedra~\cite{CousotH78} ($\poldom$);
these non-disjunctive domains all implement
the corresponding ``standard'' widening operator.
We also extended Apron by adding a prototype implementation of
the finite powerset $\psetdom = \fndom{\poldom}$
of convex polyhedra~\cite{BagnaraHZ06};
since this prototype is not (yet) provided with a widening operator,
it can only be used in the descending phase of the analysis.
Note that PAGAI features a variant analysis technique that is meant
to compute disjunctive invariants, but this would yield an analysis
which is quite different from the direct adoption of a powerset
domain; for instance, one would be forced to choose
in advance the maximal number of disjuncts that are allowed.

The experimental evaluation considers 35 C source files distributed
with PAGAI, which are variants of benchmarks taken from the SNU real-time
benchmark suite for worst-case execution time analysis.
PAGAI can be configured to perform a precision comparison
among two different abstract domains (DOM$_1$ and DOM$_2$):
in this case, the analyzer records
the invariant properties computed by the two domains
for each widening point (WP);
then it compares them and provides a final report
made of four numbers, counting the widening points on which
the invariant computed by the first domain is, respectively,
equivalent (EQ), stronger (LT), weaker (GT) and uncomparable (UN)
with respect to the invariant computed by the second domain.
The results of the precision comparisons have been summarized
in Tables~\ref{tab:prec-bop} and~\ref{tab:prec-pset};
note that, for readability, the tables show the percentages
of widening points, rather than absolute values.%
\footnote{The total number of widening points is 281;
detailed tables are available in the appendix.}


\begin{table}
\begin{centering}
\begin{tabular}{|c|c|rrrr|r|}
\hhline{~~----~}
\multicolumn{2}{c|}{} & \multicolumn{4}{c|}{\% WP} \\
\hline
 DOM$_1$ & DOM$_2$ & EQ & LT & GT & UN & $\Delta\mathrm{EQ}$ \\
\hline
$\boxdom$ & $\octdom$
  & \textbf{66.5} & 0.7 & 32.4 & 0.4 & \\
$\boxdom$ & \ascdsc{$\boxdom$}{$\octdom$}
  & 83.6 & 0.4 & \textbf{16.0} & 0.0 & \\
\ascdsc{$\boxdom$}{$\octdom$} & $\octdom$
  & \textbf{73.0} & 0.7 & 26.3 & 0.0 & \textbf{6.4} \\
\hline
$\boxdom$ & $\poldom$
  & \textbf{53.7} & 3.6 & 37.0 & 5.7 & \\
$\boxdom$ & \ascdsc{$\boxdom$}{$\poldom$}
  & 76.5 & 0.4 & \textbf{23.1} & 0.0 & \\
\ascdsc{$\boxdom$}{$\poldom$} & $\poldom$
  & \textbf{59.8} & 5.7 & 31.3 & 3.2 & \textbf{6.0} \\
\hline
$\octdom$ & $\poldom$
  & \textbf{69.4} & 6.8 & 21.4 & 2.5 & \\
$\octdom$ & \ascdsc{$\octdom$}{$\poldom$}
  & 87.2 & 0.0 & \textbf{12.8} & 0.0 & \\
\ascdsc{$\octdom$}{$\poldom$} & $\poldom$
  & \textbf{72.2} & 8.2 & 18.5 & 1.1 & \textbf{2.8} \\
\hline
\end{tabular}
\caption{Precision comparison for non-disjunctive domains.}
\label{tab:prec-bop}
\end{centering}
\end{table}

Consider first Table~\ref{tab:prec-bop}, which is meant to
evaluate the effectiveness of the new approach when
both abstract domains are non-disjunctive.
Note that the rows in the table are divided in three groups
(three rows per group); let us focus on the first group,
which is evaluating the precision improvements obtained
when using the abstract domain \ascdsc{$\boxdom$}{$\octdom$}.
The first row in the group provides the \emph{baseline}
for the precision comparison:
in particular, the value in column EQ (highlighted in boldface)
informs us that the domain $\boxdom$ achieves the same precision
as $\octdom$ on 66.5\% of the widening points; this means that
only the remaining 33.5\% of widening points are further \emph{improvable}.
The second row in the group, in particular the value in column GT,
shows us that \ascdsc{$\boxdom$}{$\octdom$} is able to improve the
precision of $\boxdom$ on 16\% of all widening points.
The third row in the group, in particular the value in column EQ,
informs us that \ascdsc{$\boxdom$}{$\octdom$}
is able to achieve the same precision of $\octdom$ on 73\% of the
widening points: this corresponds to an increase by 6.4\%
(reported in the column labeled $\Delta\mathrm{EQ}$)
with respect to the baseline EQ value (in the first row).

It is worth stressing that the percentages highlighted in the
second and third row of the group are computed with respect to
the total number of widening points, which might mislead the
reader towards an underestimation of the effectiveness of the approach.
One should observe that a precision gain on 16.0\% of \emph{all}
the widening points corresponds to
a precision gain on almost one half ($16.0 / 33.5 = 47.9\%$)
of the \emph{improvable} widening points.
The same reasoning applies to the 6.4\% value of $\Delta\mathrm{EQ}$,
which corresponds to almost 20\% of the improvable widening points.

Similar observations can be derived from the second and third group
of rows in Table~\ref{tab:prec-bop},
where we evaluate the abstract domain combinations
\ascdsc{$\boxdom$}{$\poldom$} and \ascdsc{$\octdom$}{$\poldom$},
respectively.
For instance, the third group of rows in Table~\ref{tab:prec-bop}
informs us that \ascdsc{$\octdom$}{$\poldom$} is able to improve
precision on 12.8\% of all the widening points with respect to $\octdom$
and that it increases by 2.8\% the percentage of widening points
on which the same precision as $\poldom$ is obtained.



\begin{table}
\begin{centering}
\begin{tabular}{|c|c|rrrr|rr|}
\hhline{~~------}
\multicolumn{2}{c|}{}
  & \multicolumn{4}{c|}{\% WP} & \multicolumn{2}{c|}{Time (s)} \\
\hline
 DOM$_1$ & DOM$_2$ & EQ & LT & GT & UN & DOM$_1$ & DOM$_2$ \\
\hline
$\boxdom$ & \ascdsc{$\boxdom$}{$\psetdom$}
  & 52.3 & 0.4 & \textbf{47.3} & 0.0
  & 6.20 & 6.76 \\
$\octdom$ & \ascdsc{$\octdom$}{$\psetdom$}
  & 56.2 & 0.0 & \textbf{43.8} & 0.0
  & 12.70 & 8.93 \\
$\poldom$ & \ascdsc{$\poldom$}{$\psetdom$}
  & 64.8 & 0.0 & \textbf{35.2} & 0.0
  & 7.03 & 7.53 \\
\hline
\end{tabular}
\caption{Precision comparison when using $\psetdom$ in the descending phase.}
\label{tab:prec-pset}
\end{centering}
\end{table}


In Table~\ref{tab:prec-pset} we provide the summary for
the precision comparisons between the three non-disjunctive
domains $\boxdom$, $\octdom$ and $\poldom$ and
the corresponding enhanced combinations
using the finite powerset of polyhedra $\psetdom$
in the descending phase.
Note that, in contrast with what we did in Table~\ref{tab:prec-bop},
in this case we cannot provide a \emph{baseline} comparison
with $\psetdom$ because, as said before,
this domain is missing a widening operator
and hence cannot be used in the ascending phase of the analysis.
The use of a powerset domain in the descending phase is of particular
interest because it should be able to avoid the over-approximations
that are incurred by the non-disjunctive domains when merging
control flow paths.
In fact, the values in column GT show us that the number of widening
points where a precision improvement is obtained is significantly higher
than those of Table~\ref{tab:prec-bop}, ranging from 35.2\% to 47.3\%.
In summary, the results in Tables~\ref{tab:prec-bop} and~\ref{tab:prec-pset}
provide an evidence that the adoption of a more precise abstract domain
in the descending phase of the analysis is able to significantly improve
precision. Intuitively, this is due to the fact that, by changing the
abstract domain, we are potentially improving the precision of
all the abstract semantic operators used in the descending phase
(i.e., all operators except widening).

Note that we do not perform a proper efficiency comparison, because
the considered benchmark suite seems inadequate to the purpose;
also, PAGAI is a static analyzer meant to simplify experiments,
rather than achieve maximum efficiency.
Hence, we merely report in the last two columns of
Table~\ref{tab:prec-pset} the overall time spent on the 35 tests.
A meaningful efficiency comparison will be the subject of future work.

\paragraph{A note on the relative precision of abstract domains.}
A non-expert but attentive reader may be wondering how it is possible
that the more precise abstract domain $\poldom$ can sometimes compute
weaker invariants when compared to the less precise domain $\boxdom$
(more generally, why column LT is not always zero).
A first reason is that widening operators are not monotonic;
another reason is that the two domains may be adopting different
approximation strategies for some of the semantic operators
(e.g., when modeling non-linear tests/assignments and
when taking into account the integrality of program variables).

\paragraph{A technical note on the precision comparison.}
When comparing the invariants computed
by different abstract domains, PAGAI calls a third-party model checking
tool based on SMT (Satisfiability Modulo Theory), which also takes
into account the integrality of program variables.
Hence, when comparing the abstract elements of different domains,
we are \emph{not} counting those ``dummy'' precision improvements
that are simply induced by the real relaxation step.
As a concrete example, when $x$ is an integral variable,
the $\boxdom$ value $x \in [0, 2]$,
the $\poldom$ value $\{ 0 \leq x \leq 2 \}$ and
the $\psetdom$ value $\{\;\{ x = 0 \}, \{ x = 1 \}, \{ x = 2 \}\;\}$
are all considered equivalent
(note that the last two would not be considered equivalent
when compared in the more precise domain $\psetdom$).


\subsection{A detailed example}

\begin{figure}[t]
  \centering
\begin{tabular}{c}
\begin{lstlisting}
int fib(int N) {
  int P = 0, F = 1;
  for (int K = 2; K < N; ++K) {
    /* widening point */
    int tmp = F;
    F += P;
    P = tmp;
  }
  return F;
}

int main() { return fib(7); }
\end{lstlisting}
\end{tabular}
\caption{A simplified version of \texttt{fibcall.c}}
\label{fig:fibonacci}
\end{figure}

In Figure~\ref{fig:fibonacci} we show a simplified version of
one of the tests distributed with PAGAI;
assuming $N > 1$, function \texttt{fib(N)} computes the $(N+1)$-th
element of the Fibonacci sequence $0, 1, 1, 2, 3, 5, 8, \dots$.
In Table~\ref{tab:fibcall} we show the abstract values computed for
the one and only widening point (whose position in the code
is highlighted using a comment),
first with the classical $\boxdom$ domain and
then using the \ascdsc{$\boxdom$}{$\psetdom$} combination,
i.e., using the finite powerset of polyhedra in the descending phase.
For this example, the threshold value for the number of
downward iterations is set to 10.

When using the $\boxdom$ domain,
the ascending phase ends on the 4th iteration:
due to the use of widenings, the computed post-fixpoint has
no upper bound for variable $K$; the upper bound $6$
is easily recovered in the first iteration of the descending phase,
which is then detected to be an abstract fixpoint on $\boxdom$.
When using the \ascdsc{$\boxdom$}{$\psetdom$} combination,
the ascending phase is computed exactly as before but,
before starting the descending phase,
the post-fixpoint on $\boxdom$ is transferred to the $\psetdom$ domain
using the concretization function $\fund{\gamma}{\boxdom}{\psetdom}$,
obtaining a singleton set of polyhedra (see row labeled `dsc/0').
Then the analysis proceeds by computing the descending iterates
using the more precise domain $\psetdom$;
the descending sequence is able to improve precision
by computing several disjuncts, detecting the fixpoint
on the 6th downward iteration.

\begin{small}
\begin{table}
\renewcommand{\cellalign}{tl}
\begin{tabular}{c|c|l}
domain & phase/iter & abstract value \\
\hline
$\boxdom$ & asc/1 &
  $P \in [0,0], F \in [1,1], K \in [2,2]$ \\
$\boxdom$ & asc/2 &
  $P \in [0,+\infty], F \in [1,1], K \in [2,+\infty]$ \\
$\boxdom$ & asc/3 &
  $P \in [0,+\infty], F \in [1,+\infty], K \in [2,+\infty]$ \\
$\boxdom$ & asc/4 &
  same value (detected post-fixpoint in $\boxdom$) \\
\hline
$\boxdom$ & dsc/1 &
  $P \in [0,+\infty], F \in [1,+\infty], K \in [2,6]$ \\
$\boxdom$ & dsc/2 &
  same value (detected fixpoint in $\boxdom$) \\
\hline
$\psetdom$ & dsc/0 &
  \(
  \{\;\{ P \geq 0, F \geq 1, K \geq 2 \}\:\}
  \) \\
$\psetdom$ & dsc/1 &
  \(
  \{\;\{ P = 0, F = 1, K = 2 \}, 
  \{ P \geq 1, F \geq P, 3 \leq K \leq 6 \} \;\}
  \) \\
$\psetdom$ & dsc/2 & \makecell{
  $\{\;\{ P = 0, F = 1, K = 2 \}, \{ P = 1, F = 1, K = 3 \},$ \\
  $\phantom{\{\;} \{ P+1 \leq F \leq 2P, 4 \leq K \leq 6 \} \;\}$
  } \\
$\psetdom$ & dsc/3 & \makecell{
    $\{\;\{ P = 0, F = 1, K = 2 \}, \{ P = 1, F = 1, K = 3 \},$ \\
    $\phantom{\{\;} \{ P = 1, F = 2, K = 4 \}, 
    \{ 3P \leq 2F, F \leq 2P - 1, 5 \leq K \leq 6 \}\;\}$
    } \\
$\psetdom$ & dsc/4 & \makecell{
    $\{\;\{ P = 0, F = 1, K = 2 \}, \{ P = 1, F = 1, K = 3 \}$, \\
    $\phantom{\{\;} \{ P = 1, F = 2, K = 4 \},
    \{ P = 2, F = 3, K = 5 \}$, \\
    $\phantom{\{\;} \{ 3P+1 \leq 2F, 3F \leq 5P, K = 6 \} \;\}$
    } \\
$\psetdom$ & dsc/5 & \makecell{
  $\{\; \{ P = 0, F = 1, K = 2 \}, \{ P = 1, F = 1, K = 3 \}$, \\
  $\phantom{\{\;} \{ P = 1, F = 2, K = 4 \},  \{ P = 2, F = 3, K = 5 \}$, \\
  $\phantom{\{\;} \{ P = 3, F = 5, K = 6 \} \;\}$
  } \\
$\psetdom$ & dsc/6 & same value (detected fixpoint in $\psetdom$) \\
  \hline
\end{tabular}
\caption{Abstract values computed using
  $\boxdom$ and \ascdsc{$\boxdom$}{$\psetdom$}.}
\label{tab:fibcall}
\end{table}
\end{small}


It is worth stressing that, in this specific example,
the descending sequence is able to reach a fixpoint on $\psetdom$
only because function \texttt{fib} is called with
a constant argument ($N = 7$).
If instead the value of the argument was unknown,
the descending sequence on $\psetdom$ would be non-stabilizing,
generating a new disjunct at each iteration.
This is not a real issue because, as we already said,
once started the descending phase the static analysis
can be stopped at any iteration and still preserve correctness;
a precision improvement with respect to the $\boxdom$ decreasing
sequence is obtained even when computing a single downward iterate.
Note that, for this detailed example, we have chosen
the domain combination \ascdsc{$\boxdom$}{$\psetdom$}
and the constant value $N = 7$ merely for exposition purposes,
since the computed abstract values turn out to be simpler.
For instance, if using the combination \ascdsc{$\boxdom$}{$\poldom$}
and stopping after the 3rd downward iteration, we would obtain
as post-fixpoint the following abstract value:
\begin{small}
\begin{multline*}
\{ 2 \leq K \leq 6, 3K-3P+2F \geq 8, 7K-7P-13F \leq 1,
   K+12P-8F \leq 7, \\
   K+4P-4F \leq 3, 3K-16P+8F \leq 14, 3K-6P-2F \leq 4 \}.
\end{multline*}
\end{small}

\section{Related work}\label{sect:rel}

Widening operators are quite often necessary to enforce the stabilization
of the ascending iteration sequence. Sometimes they are used
even in abstract domains having no infinite ascending chains,
to accelerate convergence, rather than enforcing it.
In restricted cases, the use of widenings can be avoided even though
the domain has infinite ascending chains:
sometimes it is possible to apply fixpoint acceleration
techniques~\cite{GonnordH06}
or strategy/policy iteration~\cite{GaubertGTZ07,GawlitzaS07}
so as to compute the \emph{exact} abstract fixpoint.

When widening operators are actually used, they also are one of
the main sources of imprecision for the static analysis.
As a consequence, many techniques try to mitigate the corresponding
precision loss:
\cite{HalbwachsPR94} proposes the \emph{widening up-to} technique,
which tries to preserve precision by using a fixed set of constraints,
used as \emph{widening hints};
a similar approach (\emph{widening with thresholds}) is used
in~\cite{BlanchetCCFMMMR03};
in~\cite{BagnaraHRZ05} a framework is proposed to improve the precision
of any given widening operator using several heuristics, while still
guaranteeing termination;
other generic techniques include
widening with landmarks~\cite{SimonK06},
lookahead widening~\cite{GopanR06},
guided static analysis~\cite{GopanR07},
and stratified widening~\cite{MonniauxG12}.
Note that all of the approaches above focus on the ascending sequence
and hence are in principle orthogonal with respect to
(i.e., they can be combined with) our proposal.

The computation of the descending sequence with narrowing is just another
technique (as a matter of fact, the very first one proposed in the
literature) to mitigate the imprecision of widenings. However,
narrowings have received fewer attention,%
\footnote{Probably, this is due to the fact that the abstract domain
glb operator implements a correct narrowing as soon as we can enforce
a finite number of applications.}
and it is often believed that the descending sequence can hardly improve
precision after a few iterations.
Such a belief is probably justified when considering abstract domains
whose elements are expressible as template polyhedra.
In particular, for the case of template polyhedra with integral bounds
(including integral boxes and octagons), \cite{AmatoMMS18} first shows that
the abstract join operation can be safely replaced by its \emph{left strict}
variant; then they prove that, when using this join operator, the
computed descending sequence cannot be infinite.
However, as witnessed by the \texttt{fibcall.c} example shown
in Figure~\ref{fig:fibonacci}, when adopting more precise domains
the descending sequence can improve the precision of the analysis
well beyond the first few iterations. This seems to be the case,
in particular, for domains such as the finite powerset of polyhedra.

\cite{AmatoS13,AmatoSSAV16} propose a technique to intertwine the
computation of widenings and narrowing (i.e., the computation of ascending
and descending chains) during the analysis, aiming at improving the
precision of the post-fixpoint computed when the CFG has nested loops.
\cite{HalbwachsH12,BoutonnetH18} propose a technique to improve the
precision of the analysis by restarting, possibly several times,
the abstract (ascending and descending) iteration sequence from
a perturbation of the computed post-fixpoint.
In the proposals recalled above the abstract domain is fixed
during the runs of the analysis, i.e., the same domain is used in the
ascending and descending iteration phases; hence, once again,
these approaches are orthogonal with respect to the proposal of this paper.
We plan to better investigate the potential synergies arising by
integrating the intertwining of widening and narrowing of~\cite{AmatoSSAV16}
(implemented for instance in IKOS~\cite{BratNSV14}
and SeaHorn~\cite{GurfinkelKKN15})
with our decoupling of the ascending and descending phases:
in practice, for the combination \ascdsc{$\ascdom$}{$\descdom$},
besides using
the concretization function $\fund{\gamma}{\ascdom}{\descdom}$
to transfer the ascending post-fixpoint to domain $\descdom$
(as in the current proposal),
we will also be using
the abstraction function $\fund{\alpha}{\descdom}{\ascdom}$
to transfer back the descending post-fixpoint whenever
restarting the ascending phase on $\ascdom$.

As mentioned previously, a formal justification for the correctness
of our proposal is easily obtained by casting it as a
meta-abstract interpretation
(the so-called A$^2$I framework~\cite{CousotGR19}).
The pre-analysis of the CFG proposed in~\cite{Bourdoncle93}
to reduce the number widening points can be interpreted as
a very early instance of the offline A$^2$I approach.
More recently, \cite{MansurMCNW20} propose an offline pre-analysis
to tailor the configuration of the static analysis tool to
the specific program being analyzed.
Online (i.e., dynamically computed) meta-analyses
include, for instance,
variable partitioning techniques~\cite{HalbwachsMG06,SinghPV17}
and the optimized implementation of semantic operators using
\emph{boxed polyhedra}~\cite{BecchiZ19}.
While there certainly are static analysis tools that perform
a \emph{non-uniform} analysis (i.e., they use different abstract domains
for different portions of the program being analyzed),
to the best of our knowledge our approach is the first example
of an analysis where the whole abstract domain (and not just one
of its operators) is changed \emph{during} the analysis of
a single portion of code.

\section{Conclusion}\label{sect:concl}
In this paper we have proposed a novel yet simple variation of the
typical approach used in static analysis by abstract interpretation,
where we decouple the ascending and descending phases
of the abstract semantics computation.
We use an abstract domain combination denoted
\ascdsc{$\ascdom$}{$\descdom$}, meaning that the ascending phase
uses an (ascending) abstract domain $\ascdom$,
while the descending phase uses a strictly more precise (descending)
abstract domain $\descdom$.
We have implemented our approach by extending the static analysis tool
PAGAI and studied its effectiveness on several, different choices
for $\ascdom$ and $\descdom$ of classical numerical abstract domains,
including boxes, octagons, convex polyhedra and sets of polyhedra.
Our preliminary experimental results show that decoupling
the ascending and descending phases in \ascdsc{$\ascdom$}{$\descdom$}
allows to obtain significant precision improvements when compared with
a classical static analysis computing on $\ascdom$.
In particular, the choice of a disjunctive domain for the
descending phase seems promising.

Even though this preliminary experimental evaluation is not adequate
for assessing the impact on efficiency (in particular, scalability)
of the proposed approach, we conjecture that the idea of using
a more precise domain $\descdom$ only in the descending phase
naturally leads to a more easily tunable efficiency/precision tradeoff.
We would also like to stress that our approach is not really meant
to be used \emph{uniformly} on all the code being analyzed; rather,
the idea is to selectively enable it on those portions of the program
where a precision gain would be desirable, but scalability issues
likely prevent to perform the whole analysis using the more precise
(and usually less efficient) domain $\descdom$.
As a consequence, an interesting problem that will be studied
in future work is how to automatically identify those parts
of the program where the decoupled approach is going to be more helpful.
In particular, we plan to investigate the effectiveness of
simple heuristics (e.g., suitable metrics on the CFG of a function)
as well as more sophisticated approaches possibly based on
machine learning techniques.
Going even further, we could not only select {\em where} to enable the more concrete descending domain $\descdom$, but also \emph{drive}
the choice of the descending domain $\descdom$. In particular, we can observe that precision of static analysis is an intensional property, namely it depends on the way the program is written \cite{BruniGGGP20,BruniGGR21}. This implies that we can drive the choice of the descending domain depending on the syntactic characteristics of expressions (guards and assignments) that, in the program, we effectively aim to analyze, since precisely these expressions are the program elements determining the precision of the analyzer \cite{BruniGGR21}.

By studying the results of the experimental evaluation, one can
also observe that, in a high percentage of cases, the analysis with
$\ascdesc{\ascdom}{\descdom}$ is able to produce the same analysis
results of the more precise domain $\descdom$
(e.g., in the 73\% of the widening points,
$\ascdesc{\boxdom}{\octdom}$ has the same results of $\octdom$ in the
considered benchmarks). This suggests an alternative usage of
the decoupled approach, starting from rather different motivations:
instead of improving the precision of a classical analysis
on $\ascdom$ using the more precise combination
$\ascdesc{\ascdom}{\descdom}$ (as discussed above),
one may try to improve the efficiency of a classical analysis
on $\descdom$ by adopting the less precise combination
$\ascdesc{\ascdom}{\descdom}$. In such a context,
one would be interested in identifying those portions of the program
where the decoupled approach is anyway as precise as the classical
approach using $\descdom$; once again, from a practical point of view,
this problem can be addressed using heuristics and/or machine learning
techniques. The same problem can also be addressed from a more
theoretical point of view, leading to the following research question:
\emph{``For a program $P$ and an abstract domain $\descdom$,
which is the less precise domain $\ascdom$ such that
the decoupled approach $\ascdesc{\ascdom}{\descdom}$
yields the same results of $\descdom$ on $P$?''}

\bibliographystyle{splncs04}
\bibliography{biblio}

\begin{thebibliography}{10}
\providecommand{\url}[1]{\texttt{#1}}
\providecommand{\urlprefix}{URL }
\providecommand{\doi}[1]{https://doi.org/#1}

\bibitem{AmatoMMS18}
Amato, G., {Di Nardo Di Maio}, S., Meo, M., Scozzari, F.: Descending chains and
  narrowing on template abstract domains. Acta Informatica  \textbf{55}(6),
  521--545 (2018). \doi{10.1007/s00236-016-0291-0}

\bibitem{AmatoS13}
Amato, G., Scozzari, F.: Localizing widening and narrowing. In: Logozzo, F.,
  F{\"{a}}hndrich, M. (eds.) Static Analysis - 20th International Symposium,
  {SAS} 2013, Seattle, WA, USA, June 20-22, 2013. Proceedings. Lecture Notes in
  Computer Science, vol.~7935, pp. 25--42. Springer (2013).
  \doi{10.1007/978-3-642-38856-9\_4}

\bibitem{AmatoSSAV16}
Amato, G., Scozzari, F., Seidl, H., Apinis, K., Vojdani, V.: Efficiently
  intertwining widening and narrowing. Sci. Comput. Program.  \textbf{120},
  1--24 (2016). \doi{10.1016/j.scico.2015.12.005}

\bibitem{Arceri20}
Arceri, V., Mastroeni, I., Xu, S.: Static analysis for ecmascript string
  manipulation programs. Appl. Sci.  \textbf{10}, ~3525 (2020).
  \doi{10.3390/app10103525}

\bibitem{BagnaraHRZ05}
Bagnara, R., Hill, P., Ricci, E., Zaffanella, E.: Precise widening operators
  for convex polyhedra. Sci. Comput. Program.  \textbf{58}(1-2),  28--56
  (2005). \doi{10.1016/j.scico.2005.02.003}

\bibitem{BagnaraHZ06}
Bagnara, R., Hill, P., Zaffanella, E.: Widening operators for powerset domains.
  Int. J. Softw. Tools Technol. Transf.  \textbf{8}(4-5),  449--466 (2006).
  \doi{10.1007/s10009-005-0215-8}

\bibitem{BecchiZ19}
Becchi, A., Zaffanella, E.: Revisiting polyhedral analysis for hybrid systems.
  In: Chang, B. (ed.) Static Analysis - 26th International Symposium, {SAS}
  2019, Porto, Portugal, October 8-11, 2019, Proceedings. Lecture Notes in
  Computer Science, vol. 11822, pp. 183--202. Springer (2019).
  \doi{10.1007/978-3-030-32304-2\_10}

\bibitem{BlanchetCCFMMMR03}
Blanchet, B., Cousot, P., Cousot, R., Feret, J., Mauborgne, L., Min{\'{e}}, A.,
  Monniaux, D., Rival, X.: A static analyzer for large safety-critical
  software. In: Cytron, R., Gupta, R. (eds.) Proceedings of the {ACM} {SIGPLAN}
  2003 Conference on Programming Language Design and Implementation 2003, San
  Diego, California, USA, June 9-11, 2003. pp. 196--207. {ACM} (2003).
  \doi{10.1145/781131.781153}

\bibitem{Bourdoncle93}
Bourdoncle, F.: Efficient chaotic iteration strategies with widenings. In:
  Bj{\o}rner, D., Broy, M., Pottosin, I. (eds.) Formal Methods in Programming
  and Their Applications, International Conference, Akademgorodok, Novosibirsk,
  Russia, June 28 - July 2, 1993, Proceedings. Lecture Notes in Computer
  Science, vol.~735, pp. 128--141. Springer (1993). \doi{10.1007/BFb0039704}

\bibitem{BoutonnetH18}
Boutonnet, R., Halbwachs, N.: Improving the results of program analysis by
  abstract interpretation beyond the decreasing sequence. Formal Methods Syst.
  Des.  \textbf{53}(3),  384--406 (2018). \doi{10.1007/s10703-017-0310-y}

\bibitem{BratNSV14}
Brat, G., Navas, J., Shi, N., Venet, A.: {IKOS:} {A} framework for static
  analysis based on abstract interpretation. In: Giannakopoulou, D.,
  Sala{\"{u}}n, G. (eds.) Software Engineering and Formal Methods - 12th
  International Conference, {SEFM} 2014, Grenoble, France, September 1-5, 2014.
  Proceedings. Lecture Notes in Computer Science, vol.~8702, pp. 271--277.
  Springer (2014). \doi{10.1007/978-3-319-10431-7\_20}

\bibitem{BruniGGGP20}
Bruni, R., Giacobazzi, R., Gori, R., Garcia{-}Contreras, I., Pavlovic, D.:
  Abstract extensionality: on the properties of incomplete abstract
  interpretations. Proc. {ACM} Program. Lang.  \textbf{4}({POPL}),  28:1--28:28
  (2020)

\bibitem{BruniGGR21}
Bruni, R., Giacobazzi, R., Gori, R., Ranzato, F.: A logic for locally complete
  abstract interpretations. In: 36th Annual {ACM/IEEE} Symposium on Logic in
  Computer Science, {LICS} 2021, Rome, Italy, June 29 - July 2, 2021. pp.
  1--13. {IEEE} (2021)

\bibitem{CC77}
Cousot, P., Cousot, R.: Abstract interpretation: {A} unified lattice model for
  static analysis of programs by construction or approximation of fixpoints.
  In: Conference Record of the Fourth {ACM} Symposium on Principles of
  Programming Languages, Los Angeles, California, USA, January 1977. pp.
  238--252 (1977)

\bibitem{CC79}
Cousot, P., Cousot, R.: Systematic design of program analysis frameworks. In:
  Conference Record of the Sixth Annual {ACM} Symposium on Principles of
  Programming Languages, San Antonio, Texas, USA, January 1979. pp. 269--282
  (1979)

\bibitem{CC92}
Cousot, P., Cousot, R.: Abstract interpretation frameworks. J. Log. Comput.
  \textbf{2}(4),  511--547 (1992). \doi{10.1093/logcom/2.4.511}

\bibitem{CousotGR19}
Cousot, P., Giacobazzi, R., Ranzato, F.: {A{\({^2}\)}I}: abstract{\({^2}\)}
  interpretation. Proc. {ACM} Program. Lang.  \textbf{3}({POPL}),  42:1--42:31
  (2019). \doi{10.1145/3290355}

\bibitem{CousotH78}
Cousot, P., Halbwachs, N.: Automatic discovery of linear restraints among
  variables of a program. In: Aho, A., Zilles, S., Szymanski, T. (eds.)
  Conference Record of the Fifth Annual {ACM} Symposium on Principles of
  Programming Languages, Tucson, Arizona, USA, January 1978. pp. 84--96. {ACM}
  Press (1978). \doi{10.1145/512760.512770}

\bibitem{CC92b}
Cousot, P., Cousot, R.: Comparing the galois connection and widening/narrowing
  approaches to abstract interpretation. In: Bruynooghe, M., Wirsing, M. (eds.)
  Programming Language Implementation and Logic Programming, 4th International
  Symposium, PLILP'92, Leuven, Belgium, August 26-28, 1992, Proceedings.
  Lecture Notes in Computer Science, vol.~631, pp. 269--295. Springer (1992).
  \doi{10.1007/3-540-55844-6\_142}

\bibitem{GaubertGTZ07}
Gaubert, S., Goubault, E., Taly, A., Zennou, S.: Static analysis by policy
  iteration on relational domains. In: Nicola, R.D. (ed.) Programming Languages
  and Systems, 16th European Symposium on Programming, {ESOP} 2007, Held as
  Part of the Joint European Conferences on Theory and Practics of Software,
  {ETAPS} 2007, Braga, Portugal, March 24 - April 1, 2007, Proceedings. Lecture
  Notes in Computer Science, vol.~4421, pp. 237--252. Springer (2007).
  \doi{10.1007/978-3-540-71316-6\_17}

\bibitem{GawlitzaS07}
Gawlitza, T., Seidl, H.: Precise fixpoint computation through strategy
  iteration. In: Nicola, R.D. (ed.) Programming Languages and Systems, 16th
  European Symposium on Programming, {ESOP} 2007, Held as Part of the Joint
  European Conferences on Theory and Practics of Software, {ETAPS} 2007, Braga,
  Portugal, March 24 - April 1, 2007, Proceedings. Lecture Notes in Computer
  Science, vol.~4421, pp. 300--315. Springer (2007).
  \doi{10.1007/978-3-540-71316-6\_21}

\bibitem{GonnordH06}
Gonnord, L., Halbwachs, N.: Combining widening and acceleration in linear
  relation analysis. In: Yi, K. (ed.) Static Analysis, 13th International
  Symposium, {SAS} 2006, Seoul, Korea, August 29-31, 2006, Proceedings. Lecture
  Notes in Computer Science, vol.~4134, pp. 144--160. Springer (2006).
  \doi{10.1007/11823230\_10}

\bibitem{GopanR06}
Gopan, D., Reps, T.: Lookahead widening. In: Ball, T., Jones, R. (eds.)
  Computer Aided Verification, 18th International Conference, {CAV} 2006,
  Seattle, WA, USA, August 17-20, 2006, Proceedings. Lecture Notes in Computer
  Science, vol.~4144, pp. 452--466. Springer (2006). \doi{10.1007/11817963\_41}

\bibitem{GopanR07}
Gopan, D., Reps, T.: Guided static analysis. In: Nielson, H., Fil{\'{e}}, G.
  (eds.) Static Analysis, 14th International Symposium, {SAS} 2007, Kongens
  Lyngby, Denmark, August 22-24, 2007, Proceedings. Lecture Notes in Computer
  Science, vol.~4634, pp. 349--365. Springer (2007).
  \doi{10.1007/978-3-540-74061-2\_22}

\bibitem{GurfinkelKKN15}
Gurfinkel, A., Kahsai, T., Komuravelli, A., Navas, J.: The seahorn verification
  framework. In: Kroening, D., Pasareanu, C. (eds.) Computer Aided Verification
  - 27th International Conference, {CAV} 2015, San Francisco, CA, USA, July
  18-24, 2015, Proceedings, Part {I}. Lecture Notes in Computer Science,
  vol.~9206, pp. 343--361. Springer (2015). \doi{10.1007/978-3-319-21690-4\_20}

\bibitem{HalbwachsH12}
Halbwachs, N., Henry, J.: When the decreasing sequence fails. In: Min{\'{e}},
  A., Schmidt, D. (eds.) Static Analysis - 19th International Symposium, {SAS}
  2012, Deauville, France, September 11-13, 2012. Proceedings. Lecture Notes in
  Computer Science, vol.~7460, pp. 198--213. Springer (2012).
  \doi{10.1007/978-3-642-33125-1\_15}

\bibitem{HalbwachsMG06}
Halbwachs, N., Merchat, D., Gonnord, L.: Some ways to reduce the space
  dimension in polyhedra computations. Formal Methods Syst. Des.
  \textbf{29}(1),  79--95 (2006). \doi{10.1007/s10703-006-0013-2}

\bibitem{HalbwachsPR94}
Halbwachs, N., Proy, Y., Raymond, P.: Verification of linear hybrid systems by
  means of convex approximations. In: Charlier, B.L. (ed.) Static Analysis,
  First International Static Analysis Symposium, SAS'94, Namur, Belgium,
  September 28-30, 1994, Proceedings. Lecture Notes in Computer Science,
  vol.~864, pp. 223--237. Springer (1994). \doi{10.1007/3-540-58485-4\_43}

\bibitem{HenryMM12}
Henry, J., Monniaux, D., Moy, M.: {PAGAI:} {A} path sensitive static analyser.
  Electron. Notes Theor. Comput. Sci.  \textbf{289},  15--25 (2012).
  \doi{10.1016/j.entcs.2012.11.003}

\bibitem{JeannetM09}
Jeannet, B., Min{\'{e}}, A.: Apron: {A} library of numerical abstract domains
  for static analysis. In: Bouajjani, A., Maler, O. (eds.) Computer Aided
  Verification, 21st International Conference, {CAV} 2009, Grenoble, France,
  June 26 - July 2, 2009. Proceedings. Lecture Notes in Computer Science,
  vol.~5643, pp. 661--667. Springer (2009). \doi{10.1007/978-3-642-02658-4\_52}

\bibitem{MansurMCNW20}
Mansur, M., Mariano, B., Christakis, M., Navas, J., W{\"{u}}stholz, V.:
  Automatically tailoring abstract interpretation to custom usage scenarios.
  In: Silva, A., Leino, K. (eds.) Computer Aided Verification - 33rd
  International Conference, {CAV} 2021, Virtual Event, July 20-23, 2021,
  Proceedings, Part {II}. Lecture Notes in Computer Science, vol. 12760, pp.
  777--800. Springer (2021). \doi{10.1007/978-3-030-81688-9\_36}

\bibitem{Mine06}
Min{\'{e}}, A.: The octagon abstract domain. High. Order Symb. Comput.
  \textbf{19}(1),  31--100 (2006). \doi{10.1007/s10990-006-8609-1}

\bibitem{MonniauxG12}
Monniaux, D., Guen, J.L.: Stratified static analysis based on variable
  dependencies. Electron. Notes Theor. Comput. Sci.  \textbf{288},  61--74
  (2012). \doi{10.1016/j.entcs.2012.10.008}

\bibitem{popa}
Nielson, F., Nielson, H., Hankin, C.: Principles of program analysis. Springer
  (1999). \doi{10.1007/978-3-662-03811-6}

\bibitem{SimonK06}
Simon, A., King, A.: Widening polyhedra with landmarks. In: Kobayashi, N. (ed.)
  Programming Languages and Systems, 4th Asian Symposium, {APLAS} 2006, Sydney,
  Australia, November 8-10, 2006, Proceedings. Lecture Notes in Computer
  Science, vol.~4279, pp. 166--182. Springer (2006). \doi{10.1007/11924661\_11}

\bibitem{SinghPV17}
Singh, G., P{\"{u}}schel, M., Vechev, M.: Fast polyhedra abstract domain. In:
  Castagna, G., Gordon, A. (eds.) Proceedings of the 44th {ACM} {SIGPLAN}
  Symposium on Principles of Programming Languages, {POPL} 2017, Paris, France,
  January 18-20, 2017. pp. 46--59. {ACM} (2017). \doi{10.1145/3009837.3009885}

\end{thebibliography}

\appendix
\section{Appendix}
\label{sect:appendix}

In this appendix we show the tables with the details of
the experimental evaluation summarized in Section~\ref{sect:expeval}.

\begin{table}
	\begin{centering}
		\begin{tabular}{|l|r|rrrr|rrrr|rrrrr|}
			\hhline{~~-------------}
			\multicolumn{2}{c|}{}
			                            & \multicolumn{4}{c|}{Box vs Oct}
			                            & \multicolumn{4}{c|}{Box vs \ascdsc{Box}{Oct}}
			                            & \multicolumn{5}{c|}{\ascdsc{Box}{Oct} vs Oct}                                                                                       \\
			\hhline{~~-------------}
			\multicolumn{2}{c|}{}
			                            & EQ                                            & LT  & GT   & UN  & EQ  & LT & GT & UN & EQ & LT & GT & UN & $\Delta\mathrm{EQ}$     \\
			\hline
			\multicolumn{1}{|r|}{\# WP} & 281
			                            & 187                                           & 2   & 91   & 1
			                            & 235                                           & 1   & 45   & 0
			                            & 205                                           & 2   & 74   & 0   & 18                                                               \\
			\multicolumn{1}{|r|}{\% WP} & 100.0
			                            & 66.5                                          & 0.7 & 32.4 & 0.4
			                            & 83.6                                          & 0.4 & 16.0 & 0.0
			                            & 73.0                                          & 0.7 & 26.3 & 0.0 & 6.4                                                              \\
			\hline
			adpcm                       & 27                                            & 27  & 0    & 0   & 0   & 27 & 0  & 0  & 0  & 27 & 0  & 0  & 0                   & 0 \\
			bs                          & 1                                             & 0   & 0    & 1   & 0   & 0  & 0  & 1  & 0  & 1  & 0  & 0  & 0                   & 1 \\
			bsort100                    & 3                                             & 2   & 0    & 1   & 0   & 3  & 0  & 0  & 0  & 2  & 0  & 1  & 0                   & 0 \\
			cnt                         & 4                                             & 3   & 0    & 1   & 0   & 3  & 0  & 1  & 0  & 4  & 0  & 0  & 0                   & 1 \\
			compress                    & 11                                            & 3   & 0    & 8   & 0   & 8  & 0  & 3  & 0  & 4  & 0  & 7  & 0                   & 1 \\
			cover                       & 3                                             & 3   & 0    & 0   & 0   & 3  & 0  & 0  & 0  & 3  & 0  & 0  & 0                   & 0 \\
			crc                         & 6                                             & 6   & 0    & 0   & 0   & 6  & 0  & 0  & 0  & 6  & 0  & 0  & 0                   & 0 \\
			decompress                  & 67                                            & 46  & 0    & 20  & 1   & 55 & 1  & 11 & 0  & 50 & 0  & 17 & 0                   & 4 \\
			duff                        & 1                                             & 1   & 0    & 0   & 0   & 1  & 0  & 0  & 0  & 1  & 0  & 0  & 0                   & 0 \\
			edn                         & 12                                            & 12  & 0    & 0   & 0   & 12 & 0  & 0  & 0  & 12 & 0  & 0  & 0                   & 0 \\
			expint                      & 3                                             & 3   & 0    & 0   & 0   & 3  & 0  & 0  & 0  & 3  & 0  & 0  & 0                   & 0 \\
			fac                         & 1                                             & 1   & 0    & 0   & 0   & 1  & 0  & 0  & 0  & 1  & 0  & 0  & 0                   & 0 \\
			fdct                        & 2                                             & 2   & 0    & 0   & 0   & 2  & 0  & 0  & 0  & 2  & 0  & 0  & 0                   & 0 \\
			fft1                        & 30                                            & 10  & 0    & 20  & 0   & 24 & 0  & 6  & 0  & 12 & 0  & 18 & 0                   & 2 \\
			fibcall                     & 1                                             & 0   & 0    & 1   & 0   & 0  & 0  & 1  & 0  & 1  & 0  & 0  & 0                   & 1 \\
			fir                         & 2                                             & 1   & 0    & 1   & 0   & 1  & 0  & 1  & 0  & 2  & 0  & 0  & 0                   & 1 \\
			insertsort                  & 2                                             & 0   & 0    & 2   & 0   & 1  & 0  & 1  & 0  & 0  & 0  & 2  & 0                   & 0 \\
			janne\_complex              & 2                                             & 0   & 0    & 2   & 0   & 0  & 0  & 2  & 0  & 2  & 0  & 0  & 0                   & 2 \\
			jfdctint                    & 3                                             & 2   & 0    & 1   & 0   & 2  & 0  & 1  & 0  & 3  & 0  & 0  & 0                   & 1 \\
			lcdnum                      & 1                                             & 1   & 0    & 0   & 0   & 1  & 0  & 0  & 0  & 1  & 0  & 0  & 0                   & 0 \\
			lms                         & 12                                            & 8   & 0    & 4   & 0   & 11 & 0  & 1  & 0  & 8  & 0  & 4  & 0                   & 0 \\
			loop                        & 1                                             & 1   & 0    & 0   & 0   & 1  & 0  & 0  & 0  & 1  & 0  & 0  & 0                   & 0 \\
			ludcmp                      & 11                                            & 5   & 0    & 6   & 0   & 8  & 0  & 3  & 0  & 6  & 0  & 5  & 0                   & 1 \\
			matmult                     & 7                                             & 6   & 0    & 1   & 0   & 6  & 0  & 1  & 0  & 7  & 0  & 0  & 0                   & 1 \\
			minver                      & 17                                            & 14  & 2    & 1   & 0   & 17 & 0  & 0  & 0  & 14 & 2  & 1  & 0                   & 0 \\
			ndes                        & 12                                            & 6   & 0    & 6   & 0   & 6  & 0  & 6  & 0  & 6  & 0  & 6  & 0                   & 0 \\
			ns                          & 4                                             & 4   & 0    & 0   & 0   & 4  & 0  & 0  & 0  & 4  & 0  & 0  & 0                   & 0 \\
			nsichneu                    & 1                                             & 1   & 0    & 0   & 0   & 1  & 0  & 0  & 0  & 1  & 0  & 0  & 0                   & 0 \\
			prime                       & 2                                             & 2   & 0    & 0   & 0   & 2  & 0  & 0  & 0  & 2  & 0  & 0  & 0                   & 0 \\
			qsort-exam                  & 6                                             & 1   & 0    & 5   & 0   & 4  & 0  & 2  & 0  & 1  & 0  & 5  & 0                   & 0 \\
			qurt                        & 3                                             & 3   & 0    & 0   & 0   & 3  & 0  & 0  & 0  & 3  & 0  & 0  & 0                   & 0 \\
			select                      & 4                                             & 1   & 0    & 3   & 0   & 4  & 0  & 0  & 0  & 1  & 0  & 3  & 0                   & 0 \\
			sqrt                        & 1                                             & 1   & 0    & 0   & 0   & 1  & 0  & 0  & 0  & 1  & 0  & 0  & 0                   & 0 \\
			st                          & 7                                             & 6   & 0    & 1   & 0   & 6  & 0  & 1  & 0  & 7  & 0  & 0  & 0                   & 1 \\
			ud                          & 11                                            & 5   & 0    & 6   & 0   & 8  & 0  & 3  & 0  & 6  & 0  & 5  & 0                   & 1 \\
			\hline
		\end{tabular}
		\caption{detail of the precision improvements obtained when using
			the \ascdsc{Box}{Oct} abstract domain combination
			(see 1st group of rows in Table~\ref{tab:prec-bop}).}
	\end{centering}
\end{table}


\begin{table}
	\begin{centering}
		\begin{tabular}{|l|r|rrrr|rrrr|rrrrr|}
			\hhline{~~-------------}
			\multicolumn{2}{c|}{}
			                            & \multicolumn{4}{c|}{Box vs Pol}
			                            & \multicolumn{4}{c|}{Box vs \ascdsc{Box}{Pol}}
			                            & \multicolumn{5}{c|}{\ascdsc{Box}{Pol} vs Pol}                                                                                       \\
			\hhline{~~-------------}
			\multicolumn{2}{c|}{}
			                            & EQ                                            & LT  & GT   & UN  & EQ  & LT & GT & UN & EQ & LT & GT & UN & $\Delta\mathrm{EQ}$     \\
			\hline
			\multicolumn{1}{|r|}{\# WP} & 281
			                            & 151                                           & 10  & 104  & 16
			                            & 215                                           & 1   & 65   & 0
			                            & 168                                           & 16  & 88   & 9   & 17                                                               \\
			\multicolumn{1}{|r|}{\% WP} & 100.0
			                            & 53.7                                          & 3.6 & 37.0 & 5.7
			                            & 76.5                                          & 0.4 & 23.1 & 0.0
			                            & 59.8                                          & 5.7 & 31.3 & 3.2 & 6.0                                                              \\
			\hline
			adpcm                       & 27                                            & 8   & 3    & 15  & 1   & 27 & 0  & 0  & 0  & 8  & 3  & 15 & 1                   & 0 \\
			bs                          & 1                                             & 0   & 0    & 1   & 0   & 0  & 0  & 1  & 0  & 1  & 0  & 0  & 0                   & 1 \\
			bsort100                    & 3                                             & 2   & 0    & 1   & 0   & 2  & 0  & 1  & 0  & 2  & 0  & 1  & 0                   & 0 \\
			cnt                         & 4                                             & 0   & 0    & 4   & 0   & 0  & 0  & 4  & 0  & 2  & 0  & 2  & 0                   & 2 \\
			compress                    & 11                                            & 1   & 3    & 1   & 6   & 7  & 0  & 4  & 0  & 1  & 3  & 1  & 6                   & 0 \\
			cover                       & 3                                             & 3   & 0    & 0   & 0   & 3  & 0  & 0  & 0  & 3  & 0  & 0  & 0                   & 0 \\
			crc                         & 6                                             & 6   & 0    & 0   & 0   & 6  & 0  & 0  & 0  & 6  & 0  & 0  & 0                   & 0 \\
			decompress                  & 67                                            & 46  & 0    & 20  & 1   & 55 & 1  & 11 & 0  & 50 & 0  & 17 & 0                   & 4 \\
			duff                        & 1                                             & 1   & 0    & 0   & 0   & 1  & 0  & 0  & 0  & 1  & 0  & 0  & 0                   & 0 \\
			edn                         & 12                                            & 12  & 0    & 0   & 0   & 12 & 0  & 0  & 0  & 12 & 0  & 0  & 0                   & 0 \\
			expint                      & 3                                             & 3   & 0    & 0   & 0   & 3  & 0  & 0  & 0  & 3  & 0  & 0  & 0                   & 0 \\
			fac                         & 1                                             & 1   & 0    & 0   & 0   & 1  & 0  & 0  & 0  & 1  & 0  & 0  & 0                   & 0 \\
			fdct                        & 2                                             & 2   & 0    & 0   & 0   & 2  & 0  & 0  & 0  & 2  & 0  & 0  & 0                   & 0 \\
			fft1                        & 30                                            & 5   & 0    & 17  & 8   & 13 & 0  & 17 & 0  & 8  & 6  & 14 & 2                   & 3 \\
			fibcall                     & 1                                             & 0   & 1    & 0   & 0   & 0  & 0  & 1  & 0  & 0  & 1  & 0  & 0                   & 0 \\
			fir                         & 2                                             & 1   & 0    & 1   & 0   & 1  & 0  & 1  & 0  & 2  & 0  & 0  & 0                   & 1 \\
			insertsort                  & 2                                             & 0   & 0    & 2   & 0   & 1  & 0  & 1  & 0  & 0  & 0  & 2  & 0                   & 0 \\
			janne\_complex              & 2                                             & 0   & 0    & 2   & 0   & 0  & 0  & 2  & 0  & 2  & 0  & 0  & 0                   & 2 \\
			jfdctint                    & 3                                             & 2   & 1    & 0   & 0   & 2  & 0  & 1  & 0  & 2  & 1  & 0  & 0                   & 0 \\
			lcdnum                      & 1                                             & 1   & 0    & 0   & 0   & 1  & 0  & 0  & 0  & 1  & 0  & 0  & 0                   & 0 \\
			lms                         & 12                                            & 6   & 0    & 6   & 0   & 9  & 0  & 3  & 0  & 6  & 0  & 6  & 0                   & 0 \\
			loop                        & 1                                             & 1   & 0    & 0   & 0   & 1  & 0  & 0  & 0  & 1  & 0  & 0  & 0                   & 0 \\
			ludcmp                      & 11                                            & 5   & 0    & 6   & 0   & 8  & 0  & 3  & 0  & 6  & 0  & 5  & 0                   & 1 \\
			matmult                     & 7                                             & 5   & 0    & 2   & 0   & 5  & 0  & 2  & 0  & 7  & 0  & 0  & 0                   & 2 \\
			minver                      & 17                                            & 14  & 2    & 1   & 0   & 16 & 0  & 1  & 0  & 14 & 2  & 1  & 0                   & 0 \\
			ndes                        & 12                                            & 6   & 0    & 6   & 0   & 6  & 0  & 6  & 0  & 6  & 0  & 6  & 0                   & 0 \\
			ns                          & 4                                             & 4   & 0    & 0   & 0   & 4  & 0  & 0  & 0  & 4  & 0  & 0  & 0                   & 0 \\
			nsichneu                    & 1                                             & 1   & 0    & 0   & 0   & 1  & 0  & 0  & 0  & 1  & 0  & 0  & 0                   & 0 \\
			prime                       & 2                                             & 2   & 0    & 0   & 0   & 2  & 0  & 0  & 0  & 2  & 0  & 0  & 0                   & 0 \\
			qsort-exam                  & 6                                             & 3   & 0    & 3   & 0   & 4  & 0  & 2  & 0  & 3  & 0  & 3  & 0                   & 0 \\
			qurt                        & 3                                             & 3   & 0    & 0   & 0   & 3  & 0  & 0  & 0  & 3  & 0  & 0  & 0                   & 0 \\
			select                      & 4                                             & 1   & 0    & 3   & 0   & 4  & 0  & 0  & 0  & 1  & 0  & 3  & 0                   & 0 \\
			sqrt                        & 1                                             & 1   & 0    & 0   & 0   & 1  & 0  & 0  & 0  & 1  & 0  & 0  & 0                   & 0 \\
			st                          & 7                                             & 0   & 0    & 7   & 0   & 6  & 0  & 1  & 0  & 0  & 0  & 7  & 0                   & 0 \\
			ud                          & 11                                            & 5   & 0    & 6   & 0   & 8  & 0  & 3  & 0  & 6  & 0  & 5  & 0                   & 1 \\
			\hline
		\end{tabular}
		\caption{detail of the precision comparison for
			the \ascdsc{Box}{Pol} combination
			(see 2nd group of rows in Table~\ref{tab:prec-bop}).}
	\end{centering}
\end{table}


\begin{table}
	\begin{centering}
		\begin{tabular}{|l|r|rrrr|rrrr|rrrrr|}
			\hhline{~~-------------}
			\multicolumn{2}{c|}{}
			                            & \multicolumn{4}{c|}{Oct vs Pol}
			                            & \multicolumn{4}{c|}{Oct vs \ascdsc{Oct}{Pol}}
			                            & \multicolumn{5}{c|}{\ascdsc{Oct}{Pol} vs Pol}                                                                                       \\
			\hhline{~~-------------}
			\multicolumn{2}{c|}{}
			                            & EQ                                            & LT  & GT   & UN  & EQ  & LT & GT & UN & EQ & LT & GT & UN & $\Delta\mathrm{EQ}$     \\
			\hline
			\multicolumn{1}{|r|}{\# WP} & 281
			                            & 195                                           & 19  & 60   & 7
			                            & 245                                           & 0   & 36   & 0
			                            & 203                                           & 23  & 52   & 3   & 8                                                                \\
			\multicolumn{1}{|r|}{\% WP} & 100.0
			                            & 69.4                                          & 6.8 & 21.4 & 2.5
			                            & 87.2                                          & 0.0 & 12.8 & 0.0
			                            & 72.2                                          & 8.2 & 18.5 & 1.1 & 2.8                                                              \\
			\hline
			adpcm                       & 27                                            & 8   & 3    & 15  & 1   & 27 & 0  & 0  & 0  & 8  & 3  & 15 & 1                   & 0 \\
			bs                          & 1                                             & 1   & 0    & 0   & 0   & 1  & 0  & 0  & 0  & 1  & 0  & 0  & 0                   & 0 \\
			bsort100                    & 3                                             & 2   & 0    & 1   & 0   & 2  & 0  & 1  & 0  & 3  & 0  & 0  & 0                   & 1 \\
			cnt                         & 4                                             & 0   & 0    & 4   & 0   & 0  & 0  & 4  & 0  & 2  & 0  & 2  & 0                   & 2 \\
			compress                    & 11                                            & 1   & 9    & 1   & 0   & 8  & 0  & 3  & 0  & 1  & 9  & 1  & 0                   & 0 \\
			cover                       & 3                                             & 3   & 0    & 0   & 0   & 3  & 0  & 0  & 0  & 3  & 0  & 0  & 0                   & 0 \\
			crc                         & 6                                             & 6   & 0    & 0   & 0   & 6  & 0  & 0  & 0  & 6  & 0  & 0  & 0                   & 0 \\
			decompress                  & 67                                            & 62  & 0    & 5   & 0   & 63 & 0  & 4  & 0  & 62 & 0  & 5  & 0                   & 0 \\
			duff                        & 1                                             & 1   & 0    & 0   & 0   & 1  & 0  & 0  & 0  & 1  & 0  & 0  & 0                   & 0 \\
			edn                         & 12                                            & 12  & 0    & 0   & 0   & 12 & 0  & 0  & 0  & 12 & 0  & 0  & 0                   & 0 \\
			expint                      & 3                                             & 3   & 0    & 0   & 0   & 3  & 0  & 0  & 0  & 3  & 0  & 0  & 0                   & 0 \\
			fac                         & 1                                             & 1   & 0    & 0   & 0   & 1  & 0  & 0  & 0  & 1  & 0  & 0  & 0                   & 0 \\
			fdct                        & 2                                             & 2   & 0    & 0   & 0   & 2  & 0  & 0  & 0  & 2  & 0  & 0  & 0                   & 0 \\
			fft1                        & 30                                            & 7   & 2    & 15  & 6   & 15 & 0  & 15 & 0  & 10 & 6  & 12 & 2                   & 3 \\
			fibcall                     & 1                                             & 0   & 1    & 0   & 0   & 0  & 0  & 1  & 0  & 0  & 1  & 0  & 0                   & 0 \\
			fir                         & 2                                             & 2   & 0    & 0   & 0   & 2  & 0  & 0  & 0  & 2  & 0  & 0  & 0                   & 0 \\
			insertsort                  & 2                                             & 1   & 0    & 1   & 0   & 2  & 0  & 0  & 0  & 1  & 0  & 1  & 0                   & 0 \\
			janne_complex               & 2                                             & 2   & 0    & 0   & 0   & 2  & 0  & 0  & 0  & 2  & 0  & 0  & 0                   & 0 \\
			jfdctint                    & 3                                             & 2   & 1    & 0   & 0   & 2  & 0  & 1  & 0  & 2  & 1  & 0  & 0                   & 0 \\
			lcdnum                      & 1                                             & 1   & 0    & 0   & 0   & 1  & 0  & 0  & 0  & 1  & 0  & 0  & 0                   & 0 \\
			lms                         & 12                                            & 6   & 0    & 6   & 0   & 9  & 0  & 3  & 0  & 6  & 0  & 6  & 0                   & 0 \\
			loop                        & 1                                             & 1   & 0    & 0   & 0   & 1  & 0  & 0  & 0  & 1  & 0  & 0  & 0                   & 0 \\
			ludcmp                      & 11                                            & 11  & 0    & 0   & 0   & 11 & 0  & 0  & 0  & 11 & 0  & 0  & 0                   & 0 \\
			matmult                     & 7                                             & 5   & 0    & 2   & 0   & 5  & 0  & 2  & 0  & 7  & 0  & 0  & 0                   & 2 \\
			minver                      & 17                                            & 17  & 0    & 0   & 0   & 17 & 0  & 0  & 0  & 17 & 0  & 0  & 0                   & 0 \\
			ndes                        & 12                                            & 12  & 0    & 0   & 0   & 12 & 0  & 0  & 0  & 12 & 0  & 0  & 0                   & 0 \\
			ns                          & 4                                             & 4   & 0    & 0   & 0   & 4  & 0  & 0  & 0  & 4  & 0  & 0  & 0                   & 0 \\
			nsichneu                    & 1                                             & 1   & 0    & 0   & 0   & 1  & 0  & 0  & 0  & 1  & 0  & 0  & 0                   & 0 \\
			prime                       & 2                                             & 2   & 0    & 0   & 0   & 2  & 0  & 0  & 0  & 2  & 0  & 0  & 0                   & 0 \\
			qsort-exam                  & 6                                             & 3   & 3    & 0   & 0   & 5  & 0  & 1  & 0  & 3  & 3  & 0  & 0                   & 0 \\
			qurt                        & 3                                             & 3   & 0    & 0   & 0   & 3  & 0  & 0  & 0  & 3  & 0  & 0  & 0                   & 0 \\
			select                      & 4                                             & 1   & 0    & 3   & 0   & 4  & 0  & 0  & 0  & 1  & 0  & 3  & 0                   & 0 \\
			sqrt                        & 1                                             & 1   & 0    & 0   & 0   & 1  & 0  & 0  & 0  & 1  & 0  & 0  & 0                   & 0 \\
			st                          & 7                                             & 0   & 0    & 7   & 0   & 6  & 0  & 1  & 0  & 0  & 0  & 7  & 0                   & 0 \\
			ud                          & 11                                            & 11  & 0    & 0   & 0   & 11 & 0  & 0  & 0  & 11 & 0  & 0  & 0                   & 0 \\
			\hline
		\end{tabular}
		\caption{detail of the precision comparison for
			the \ascdsc{Oct}{Pol} combination
			(see 3rd group of rows in Table~\ref{tab:prec-bop}).}
	\end{centering}
\end{table}


\begin{table}
	\begin{centering}
		\begin{tabular}{|l|r|rrrr|rrrr|rrrr|}
			\hhline{~~------------}
			\multicolumn{2}{c|}{}
			                            & \multicolumn{4}{c|}{Box vs \ascdsc{Box}{PSet}}
			                            & \multicolumn{4}{c|}{Oct vs \ascdsc{Oct}{PSet}}
			                            & \multicolumn{4}{c|}{Pol vs \ascdsc{Pol}{PSet}}                                                                \\
			\hhline{~~------------}
			\multicolumn{2}{c|}{}
			                            & EQ                                             & LT  & GT   & UN  & EQ & LT & GT & UN & EQ & LT & GT & UN     \\
			\hline
			\multicolumn{1}{|r|}{\# WP} & 281
			                            & 147                                            & 1   & 133  & 0
			                            & 158                                            & 0   & 123  & 0
			                            & 182                                            & 0   & 99   & 0                                               \\
			\multicolumn{1}{|r|}{\% WP} & 100.0
			                            & 52.3                                           & 0.4 & 47.3 & 0.0
			                            & 56.2                                           & 0.0 & 43.8 & 0.0
			                            & 64.8                                           & 0.0 & 35.2 & 0.0                                             \\
			\hline
			adpcm                       & 27                                             & 18  & 0    & 9   & 0  & 18 & 0  & 9  & 0  & 16 & 0  & 11 & 0 \\
			bs                          & 1                                              & 0   & 0    & 1   & 0  & 1  & 0  & 0  & 0  & 1  & 0  & 0  & 0 \\
			bsort100                    & 3                                              & 2   & 0    & 1   & 0  & 2  & 0  & 1  & 0  & 3  & 0  & 0  & 0 \\
			cnt                         & 4                                              & 0   & 0    & 4   & 0  & 0  & 0  & 4  & 0  & 1  & 0  & 3  & 0 \\
			compress                    & 11                                             & 5   & 0    & 6   & 0  & 5  & 0  & 6  & 0  & 8  & 0  & 3  & 0 \\
			cover                       & 3                                              & 3   & 0    & 0   & 0  & 3  & 0  & 0  & 0  & 3  & 0  & 0  & 0 \\
			crc                         & 6                                              & 5   & 0    & 1   & 0  & 5  & 0  & 1  & 0  & 5  & 0  & 1  & 0 \\
			decompress                  & 67                                             & 53  & 1    & 13  & 0  & 58 & 0  & 9  & 0  & 58 & 0  & 9  & 0 \\
			duff                        & 1                                              & 1   & 0    & 0   & 0  & 1  & 0  & 0  & 0  & 1  & 0  & 0  & 0 \\
			edn                         & 12                                             & 9   & 0    & 3   & 0  & 9  & 0  & 3  & 0  & 9  & 0  & 3  & 0 \\
			expint                      & 3                                              & 1   & 0    & 2   & 0  & 1  & 0  & 2  & 0  & 1  & 0  & 2  & 0 \\
			fac                         & 1                                              & 0   & 0    & 1   & 0  & 0  & 0  & 1  & 0  & 0  & 0  & 1  & 0 \\
			fdct                        & 2                                              & 2   & 0    & 0   & 0  & 2  & 0  & 0  & 0  & 2  & 0  & 0  & 0 \\
			fft1                        & 30                                             & 10  & 0    & 20  & 0  & 10 & 0  & 20 & 0  & 21 & 0  & 9  & 0 \\
			fibcall                     & 1                                              & 0   & 0    & 1   & 0  & 0  & 0  & 1  & 0  & 1  & 0  & 0  & 0 \\
			fir                         & 2                                              & 1   & 0    & 1   & 0  & 1  & 0  & 1  & 0  & 1  & 0  & 1  & 0 \\
			insertsort                  & 2                                              & 0   & 0    & 2   & 0  & 0  & 0  & 2  & 0  & 0  & 0  & 2  & 0 \\
			janne\_complex              & 2                                              & 0   & 0    & 2   & 0  & 1  & 0  & 1  & 0  & 1  & 0  & 1  & 0 \\
			jfdctint                    & 3                                              & 2   & 0    & 1   & 0  & 2  & 0  & 1  & 0  & 3  & 0  & 0  & 0 \\
			lcdnum                      & 1                                              & 1   & 0    & 0   & 0  & 1  & 0  & 0  & 0  & 1  & 0  & 0  & 0 \\
			lms                         & 12                                             & 4   & 0    & 8   & 0  & 4  & 0  & 8  & 0  & 5  & 0  & 7  & 0 \\
			loop                        & 1                                              & 0   & 0    & 1   & 0  & 0  & 0  & 1  & 0  & 0  & 0  & 1  & 0 \\
			ludcmp                      & 11                                             & 5   & 0    & 6   & 0  & 6  & 0  & 5  & 0  & 6  & 0  & 5  & 0 \\
			matmult                     & 7                                              & 2   & 0    & 5   & 0  & 2  & 0  & 5  & 0  & 3  & 0  & 4  & 0 \\
			minver                      & 17                                             & 8   & 0    & 9   & 0  & 8  & 0  & 9  & 0  & 8  & 0  & 9  & 0 \\
			ndes                        & 12                                             & 2   & 0    & 10  & 0  & 7  & 0  & 5  & 0  & 7  & 0  & 5  & 0 \\
			ns                          & 4                                              & 1   & 0    & 3   & 0  & 1  & 0  & 3  & 0  & 1  & 0  & 3  & 0 \\
			nsichneu                    & 1                                              & 1   & 0    & 0   & 0  & 1  & 0  & 0  & 0  & 1  & 0  & 0  & 0 \\
			prime                       & 2                                              & 0   & 0    & 2   & 0  & 0  & 0  & 2  & 0  & 0  & 0  & 2  & 0 \\
			qsort-exam                  & 6                                              & 3   & 0    & 3   & 0  & 3  & 0  & 3  & 0  & 6  & 0  & 0  & 0 \\
			qurt                        & 3                                              & 0   & 0    & 3   & 0  & 0  & 0  & 3  & 0  & 0  & 0  & 3  & 0 \\
			select                      & 4                                              & 3   & 0    & 1   & 0  & 0  & 0  & 4  & 0  & 1  & 0  & 3  & 0 \\
			sqrt                        & 1                                              & 0   & 0    & 1   & 0  & 0  & 0  & 1  & 0  & 0  & 0  & 1  & 0 \\
			st                          & 7                                              & 0   & 0    & 7   & 0  & 0  & 0  & 7  & 0  & 2  & 0  & 5  & 0 \\
			ud                          & 11                                             & 5   & 0    & 6   & 0  & 6  & 0  & 5  & 0  & 6  & 0  & 5  & 0 \\
			\hline
		\end{tabular}
		\caption{detail of the precision improvements obtained when using
			the polyhedra powerset domain PSet in the descending phase
			(see Table~\ref{tab:prec-pset}).}
	\end{centering}
\end{table}

\end{document}